\documentclass[fleqn,usenatbib]{mnras}

\usepackage{newtxtext,newtxmath}
\DeclareUnicodeCharacter{2212}{-}

\usepackage[T1]{fontenc}

\DeclareRobustCommand{\VAN}[3]{#2}
\let\VANthebibliography\thebibliography
\def\thebibliography{\DeclareRobustCommand{\VAN}[3]{##3}\VANthebibliography}

\usepackage{graphicx}	
\usepackage{amsmath}	
\usepackage{anyfontsize}
\usepackage{textcomp}   
\usepackage{soul}

\title[Galactic BPASS populations]{Gravitational wave energy spectral density properties from BPASS Galactic binary population in the Milky Way galaxy}

\author[P. Tang et al.]{%
    P. Tang$^{1,2}$\thanks{E-mail: petra.tang@auckland.ac.nz},
    R. Meyer$^{2}$,
    J.J. Eldridge$^{1}$\\
    $^{1}$Department of Physics, University of Auckland, Private Bag 92019, Auckland, New Zealand\\
    $^{2}$Department of Statistics, University of Auckland, Private Bag 92019, Auckland, New Zealand
}

\date{Accepted XXX. Received YYY; in original form ZZZ}

\pubyear{2024}

\begin{document}
\label{firstpage}
\pagerange{\pageref{firstpage}--\pageref{lastpage}}
\maketitle

\begin{abstract}
We analyse the energy spectral density properties of Gravitational waves from Galactic binary populations in the~\text{mHz} band targeted by the Laser Interferometer Space Antenna mission. Our analysis is based on combining BPASS with a Milky Way analogue galaxy from the Feedback In Realistic Environment (FIRE) simulations and the GWs these populations emit. Our investigation compares different functional forms of gravitational wave (GW) ESDs, namely the single power-law, broken power-law, and single-peak models, revealing disparities within and among Galactic binary populations. We estimate the ESDs for six different Galactic binary populations and the ESD of the total Galactic binary population for LISA. Employing a single power-law model, we predict a total Galactic binary GW signal amplitude $\alpha$ = $2.0^{+0.2}_{-0.2} \times 10^{-8}$ and a slope $\beta$ = $-2.64 ^{+0.03}_{-0.04}$ and the ESD $\rm h^2 \Omega_{GW}$ = $1.1 ^{+0.1}_{-0.1} \times 10^{-9}$ at 3~\text{mHz}. For the Galactic WDB binary GW signal $\alpha = 1^{+0.02}_{-0.02} \times 10^{-10}$, $\beta = -1.56 ^{+0.03}_{-0.03}$ and $\rm h^2 \Omega_{GW} = 18 ^{+1}_{-1} \times 10^{-12}$. Our analysis underscores the importance of accurate noise parameter estimation and highlights the complexities of modelling realistic observations, prompting future exploration into more flexible models.  

\end{abstract}
\section{Introduction}
Gravitational waves (GWs), ripples in spacetime resulting from the acceleration or merging of massive objects like black holes (BHs), neutron stars (NSs) and white dwarfs (WDs), have become a fascinating area of scientific study. These waves span a wide range of frequencies from kHz to pHz (milliseconds to billions of years), and some can only be observed from space \citep[i.e.][]{2016PhRvL.116f1102A,2017arXiv170200786A}. To further our understanding of these cosmic phenomena, the European Space Agency (ESA) is embarking on an important mission called the Laser Interferometer Space Antenna (LISA) mission, which aims to be the first space-based gravitational wave observatory (scheduled to be launched in 2035) \citep[detailed in][]{2017arXiv170200786A}.

LISA is designed to detect GW signals in the frequency ranges from 0.1~mHz to 0.1~Hz. While the scientific community eagerly awaits its deployment, we have turned to simulations and predictive modelling to gain a glimpse of the signals LISA will eventually observe. By simulating GW progenitors and predicting the GW signals that LISA is expected to detect, we can lay the groundwork for future discoveries and advance our understanding of the cosmos. In particular, LISA will be able to detect Galactic binaries and resolve many individual ultra-compact binaries \citep[UCBs, e.g. ][]{2023LRR....26....2A}. These resolved UCBs mainly consist of white dwarf binaries \citep[WDBs, e.g.][]{2004MNRAS.349..181N, 2012ApJ...758..131N, 2017MNRAS.470.1894K, 2017ApJ...846...95K, 2019MNRAS.490.5888L, 2020ApJ...898...71B, 2020ApJ...898..133L, 2022MNRAS.511.5936K, 2024arXiv240520484T}, a smaller number of neutron stars and stellar black hole binaries \citep[BHBs, e.g.][]{1990ApJ...360...75H, 2001A&A...375..890N, 2004MNRAS.349..181N, 2018MNRAS.480.2704L, 2020ApJ...892L...9A, 2020MNRAS.492.3061L, 2022ApJ...937..118W, 2024arXiv240520484T}, massive binary black holes (MBHBs) and extreme mass ratio inspirals (EMRIs) \citep[i.e.][]{2023LRR....26....2A}. Our recent study \citet{2024arXiv240520484T} predicted that, on average, four Galactic BHBs and 673 WDBs with a signal-to-noise-ratio (SNR) greater than 7 within a four-year mission will be resolved by LISA. 

There are many types of objects that LISA can detect, and neutron star binaries (NSBs) and black hole-white dwarfs (BHWDs) are rarer than neutron star-white dwarfs (BHWDs) and WDBs for LISA \citet{2017arXiv170200786A}. Based on the ATNF Pulsar Catalogue \citep[detailed in][]{2005AJ....129.1993M}, there are about 120 tight orbit NSWDs in the Milky Way disc; among them, LISA will be able to detect those systems that are close to us. One study by \citet{2018PhRvL.121m1105T} suggests that LISA could detect 50 such NSWD systems. As for the BHWD systems, with no current confirmed detection, LISA is expected to detect those with tight orbits \citep[][]{2023LRR....26....2A}. An earlier study by \citet{2001A&A...375..890N} predicts about 100 BHWD systems detectable by LISA.

However, not all binaries with an SNR greater than 7, even though detectable, will not be individually resolvable if they fall within the same frequency bin. All unresolvable binaries from the different Galactic binary populations will combine to form components of the so called ``confusion" noise and taken together will  form the Galactic stochastic GW background (SGWB). The main objective of this study is to characterise the collective effects of numerous weak, independent, and unresolved sources from different Galactic binary populations, i.e., their SGWBs \citep[][]{2017LRR....20....2R}. ``Stochastic'' implies that the SGWB can only be characterised statistically as the combined signal averages to zero and is characterised by its power per frequency bin. ``Independent'' means that each GW event is assumed to occur far away from other events, thus without interaction. ``Unresolved'' conveys that individual GW events cannot be distinguished; these unresolved signals can either be ``weak'' or strong signals overlapping in frequency and, therefore, cannot be resolved. We  assume that all resolvable signals have been subtracted from the data stream, and the focus is on the aggregate signals coming from the Galactic binary populations, not the extra-galactic astrophysical SGWB or the cosmological SGWB. Therefore, the overall characteristics of the SGWB encompass the cumulative effect of all our simulated unresolved Galactic binary populations.

In this work, we generate our Galactic binary populations in a MW-like galaxy using the Binary Population and Spectral Synthesis (BPASS\footnote{\url{http://bpass.auckland.ac.nz} or \url{http://warwick.ac.uk/bpass}}) code, tailored with constraints on physical parameters, such as mass, age, orbital period and metallicity, where all parameters are informed by observational evidence \citep[details see][]{2017PASA...34...58E,2018MNRAS.479...75S}. We do not consider each individual source but study the population as a whole and the signal it produces. This stochastic signal is characterised by its energy spectral density (ESD) per logarithmic interval of frequency which in turn can be parametrized, e.g., as in Section~\ref{sec:3}. We use a Bayesian approach to estimate the ESD parameters based on the generated galactic binary signals embedded in LISA instrument noise. We can then study these parameters, such as for instance  the amplitude and slope associated with a power law ESD, and how the parameters differ amongst different Galactic binary populations. The ESD is crucial in GW astronomy as it provides a way to understand the complete signal from compact binaries observed by LISA, enabling insight into the underlying physics of these populations without the need for detailed analysis of individual systems. By comparing predicted ESDs with LISA’s sensitivity curve, we can assess the detectability of different Galactic binary populations and determine which signals are likely to be resolved or remain in the confusion foreground. Since different binary populations likely exhibit distinct ESD signatures, analysing these spectra allows us to constrain key physical properties, such as orbital periods, mass distributions, and binary formation channels. This makes ESD estimation a powerful tool for testing population synthesis models and improving our understanding of compact binary evolution in the Milky Way. Furthermore, the Galactic foreground is an additional noise source, over an above the instrumental noise component, that needs to be taken into account when extracting point sources. For an accurate and unbiased estimation of waveform parameters of these point sources, a concise characterisation and estimation of the Galactic foreground ESD is very important \citep[][]{2025PhRvD.111b3025C}. 

In this study, we estimate the ESDs for six different Galactic binary populations and the ESD of the total Galactic binary population for LISA. This study  aids in interpreting the formation and evolution of binary systems, their role in galaxy evolution, and the formation of large-scale structures. Moreover, our models of the stochastic background originating from Galactic binary populations can \st{help to} enhance the LISA sensitivity curve. Accurately taking the galactic foreground into account will be essential to improve the detectability of the cosmological SGWB and its separability from instrumental noise \citep[][]{2021MNRAS.508..803B}. Furthermore, our analysis can be extended to accommodate cyclostationarities to improve the detectability of MW satellites \citep[i.e.][]{2025PhRvD.111f3005P}. Our findings on the ESD of GWs from Galactic binary populations can be extrapolated to other Milky-way like galaxies, potentially improving the understanding of GW signals from other galaxies. 

The collaborative utilisation of LISA data alongside data from ground-based GW detectors and electromagnetic observations spanning various wavelengths, including infrared, radio, X-ray, and gamma-rays, will significantly amplify its impact on astrophysics \citep[i.e.][]{2020PhRvD.102h4056M}. Virtually all of LISA's individually resolvable sources possess potential electromagnetic counterparts. The primary objectives moving forward involve quantifying these counterparts, evaluating the feasibility of their detection across different wavebands, and assessing their detectability relative to the corresponding GW signal's inspiral and merger stages \citep[detailed in][]{2023LRR....26....2A}. It is  crucial to evaluate this domain's current state of knowledge, particularly concerning the characteristics of stellar populations that give rise to LISA sources, such as binary evolution, stellar remnants, and their distribution within the Milky Way. Understanding the distribution and evolution of these stellar populations will enhance predictions of source properties and detection rates, contributing to a more comprehensive understanding of LISA Galactic sources.

This article is organised as follows: we introduce our simulated Galactic binary populations in Section~\ref{sec:2.1} and simulated modulated signals in Section~\ref{sec:2.2}. We then outline the formalism used in our models to generate frequency-domain noise and GW signals in Sections~\ref{sec:3.1} and \ref{sec:3.2}, respectively. For the data analysis in Section~\ref{sec:4}, we explain our Bayesian models and show the results of applying our Bayesian parametric models to BPASS Galactic binary populations in Section~\ref{sec:5}. Finally, we discuss and conclude our findings in Section~\ref{sec:6}.

\section{SGWB from BPASS Galactic binary populations}
\label{sec:2}

Various methods can be employed to simulate or generate GW signals produced by binary systems within the Milky Way galaxy. These methods often draw from comprehensive models and astrophysical simulations that incorporate different types of stellar evolution and binary interactions. One notable example is the data generation approach used in the LISA Data Challenges (LDC\footnote{\url{https://lisaldc.lal.in2p3.fr}}), which facilitates the creation of synthetic datasets for evaluating the performance of GW detection algorithms \citep[][]{2006AAS...209.7414V,2006gr.qc.....6089B, 2008APS..APRS10001V, 2010PhRvD..81f3008B}. 

In this study, we create our own simulated Galactic binary populations using BPASS and cosmological simulations with the Feedback In Realistic Environments (FIRE\footnote{\url{https://fire.northwestern.edu/}}) m12i galaxy from the Latte suite of FIRE-2 simulations \citep{2014MNRAS.445..581H, 2016ApJ...827L..23W, 2018MNRAS.480..800H}. The BPASS binary populations have been validated against observational constraints, including the rate of type Ia supernovae \citep[e.g.,][]{2008MNRAS.384.1109E, 2017PASA...34...58E, 2018MNRAS.479...75S, 2022ARA&A..60..455E}. This provides a quantitative understanding of their physical parameters and the conditions under which the maximum ESD occurs \citep[][]{2023MNRAS.524.2836V}.

The BPASS code comprises a suite of computational tools designed to model the intricate evolution of interacting binary star systems. These tools are harnessed to create synthetic stellar populations, enabling diverse predictions, including spectral energy distributions for individual stars, total stellar populations, electromagnetic and gravitational transients, and more \citep[see][and references therein]{2017PASA...34...58E, 2018MNRAS.479...75S, 2022MNRAS.514.1315B}. The underlying stellar evolution models encompass fundamental physical processes such as mass transfer (MT), stellar winds, and supernova explosions. We use the same code as our recent work in \citet{2024arXiv240520484T} to simulate the Galactic binary populations used in this study.

What sets BPASS apart is its distinctive approach to evolving stellar models with a tailored version of the Cambridge STARS code first described in \citet{2008MNRAS.384.1109E}, while the Cambridge STARS code was first presented in \citet{eggleton1971}. This approach provides a detailed representation of how stellar structures respond to binary interactions. Consequently, our models exhibit more stable mass transfer dynamics than the assumptions made in other population synthesis codes \citep[such as][and references within]{2023MNRAS.520.5724B} and compared to other binary population synthesis codes \citep[such as][]{2024arXiv240520484T}. BPASS has undergone extensive testing against a wide array of observational data, and this testing continues to expand. For instance, take the study by \citet{2023ApJ...943L..12K}, which delves into the influence of binary interactions on the chemical enrichment of galaxies, hence testing the evidence of how chemistry changes with the binary evolution.

We employ the results obtained from BPASS v2.2.1. The Galactic populations have been computed using the GW population synthesis supplementary code \textsc{Tui} (v2.jje), a GW binary population synthesis program described in \citet{2022MNRAS.511.1201G, 2023NatAs...7..444S, 2023MNRAS.520.5724B, 2023MNRAS.524.2836V}. The predictions are made using the fiducial BPASS initial binary parameters and initial mass functions. Further details regarding the BPASS code and how we combine it with a galaxy simulation can be found in our recent study \citep[see details in][]{2024arXiv240520484T}. We aim to estimate the spectral properties of Galactic binary populations based on LISA observations. 

The BPASS stellar evolution models are described in detail in \citet{2017PASA...34...58E, 2018MNRAS.479...75S}. The process of using BPASS models to create synthetic LISA populations is described in \citet{2024arXiv240520484T}. However, we present a short summary of key details here. The models assume a Kroupa initial mass function \citep{2001MNRAS.322..231K} and initial binary parameters based on observed binary populations \citep{2017ApJS..230...15M}. The initial binary parameters have all massive stars above 16~M$_{\odot}$ in binary systems with the binary fraction then dropping to approximately 20 per cent for 1~M$_{\odot}$ stars.

The stellar models are calculated by a detailed stellar evolution code that solves for the stellar structure and accounts for binary evolution. The processes included in the binary evolution include mass transfer, common envelope evolution, and supernova kicks. We note the common-envelope prescription used in BPASS models assumed angular momentum conservation \citep[the $\gamma$ prescription,][]{2020cee..book..I} rather than the more widely used energy conservation assumption \cite[the $\alpha$ prescription,][]{2020cee..book..I}. This means that in BPASS common-envelope evolution  ends with wider orbits than other binary evolution simulations.

The BPASS stellar models  cover a range of metallicity mass fractions ($Z=10^{-5}$ to $Z=0.040$). These stellar models are then combined to create synthetic stellar populations, enabling diverse predictions, including spectral energy distributions for individual stars, total stellar populations, electromagnetic and gravitational transients, and more \citep[see][and references therein]{2017PASA...34...58E, 2018MNRAS.479...75S, 2022MNRAS.514.1315B}. 

In contrast to our recent work \citep[][]{2024arXiv240520484T}, here we have expanded the different binary populations we simulate to comprise compact neutron star binaries and we consider mixtures of populations of WD and NS binaries, i.e., NSWD and BHWD systems.  This allows the full range of possible compact binaries to be included.

\subsection{BPASS Galactic binary populations}
\label{sec:2.1}

Binary population synthesis codes SeBa \citep[][]{2001A&A...375..890N} and BSE \citep[][]{2018MNRAS.480.2704L, 2019MNRAS.490.5888L} differ from BPASS in so far as they  both provide a {\em rapid} binary population synthesis. This means that stellar evolution is approximated using equations fitted to single star evolution tracks rather than being followed in detail. Consequently, the response of stars to binary interactions is more approximate in SeBa and BSE compared to BPASS, which models the full stellar structure. Additionally, SeBa and BSE assume energy conservation during common-envelope evolution, whereas BPASS assumes angular momentum conservation. As described in \citet{2024arXiv240520484T}, which compared BPASS and BSE results in detail, these differences lead to a higher occurrence of stable mass transfer in BPASS populations and a wider final period distribution for compact binaries.   Though it is not the focus of this study, we are excited that work is ongoing by the LISA Synthetic UCB Catalogue Group \citep{2023arXiv231103431V}, which aims to compare the Galactic WDB populations from several current binary population synthesis codes.

The unresolved number of Galactic binary systems within each binary population predicted from BPASS, BSE \citep[][]{2018MNRAS.480.2704L, 2019MNRAS.490.5888L}, early prediction from SeBa \citep[][]{2001A&A...375..890N} is presented in Table~\ref{tab:numbersource}, sources with signal-to-noise ratios larger then 7 have been removed. The total number of Galactic binary systems within the LISA frequency range predicted by BSE stems from personal communications with the author. The full frequency range of the predicted populations from SeBa and BPASS ranging from $10^{-10}$ to 0.1~Hz is given in the second and third column. Noticeably, only a tiny percentage of Galactic binary systems occupy the LISA frequency range of 0.1~mHz to 0.1~Hz. Of these mHz frequency sources, we only expect LISA to revolve a small number individually, and all unresolvable systems contribute to the overall Galactic SGWB. We find that both BPASS and SeBa predict a similar number of Galactic NSB, BHWD and BHNS systems; BPASS predicts more Galactic WDB and NSWD systems than SeBa; noticeably, BPASS predict less BHB than SeBa. It is also worth mentioning that in a recent paper by \citet{2024MNRAS.530..844K}, who used Seba to predict approximately $\mathcal{O}(10^2)$ detectable NSWD binaries by LISA. Please note that we do not include a comparison of ESDs between Seba and BPASS in later sections because we do not have the corresponding data. We can only compare the predicted total number of Galactic binary systems with BSE and SeBa as the predictions of other binary population synthesis of these numbers are rare. 

Table~\ref{tab:numbersource} also shows a comparison between the Galactic BHB and WDB populations and previous studies by \citet{2018MNRAS.480.2704L, 2019MNRAS.490.5888L}. Where we combined our BPASS populations and the FIRE simulations, and used \textsc{PhenomA} and \textsc{LEGWORK} to predict the populations of BPASS WDBs and BHBs in our Galaxy. In addition, we compared to earlier results using the BSE code. These differences arise from varying physical assumptions in binary evolution models. Our findings suggest that LISA's detection of compact binaries will tightly constrain binary star modeling.  A detailed discussion on the Galactic BHB and WDB population comparison is available in \citet{2024arXiv240520484T}. 

\begin{table}
  \centering
  \caption{ Number of Galactic binary systems in the MW-like galaxy predicted by two models in the LISA frequency range of greater than 0.1~mHz. The full predicted population from SeBa and BPASS of frequency ranging from $10^{-10}$ Hz to 0.1~Hz is given in the second and third columns.}
    \begin{tabular}{r|rrrr}
       Events   & SeBa     &     BPASS            & BPASS LISA         & BSE LISA       \\
          \hline
          BHB   & $2.8 \times 10^6$ &  $4.9 \times 10^5$   & 80                 & 30        \\    
          WDB   & $1.1 \times 10^8$ &  $2.7 \times 10^8$   & $1.7 \times 10^6$  & $7.7 \times 10^6$\\
          NSB   & $7.5 \times 10^5$ &  $7.0 \times 10^5$   & $7.6 \times 10^4$  & --        \\
         NSWD   & $2.2 \times 10^6$ &  $6.5 \times 10^6$   & $2.5 \times 10^5$  & --        \\
         BHNS   & $4.7 \times 10^5$ &  $3.8 \times 10^5$   & $1.1 \times 10^3$  & --        \\
         BHWD   & $1.4 \times 10^6$ &  $2.4 \times 10^6$   & $5.2 \times 10^4$  & --        \\ 
    \end{tabular}%
  \label{tab:numbersource}%
\end{table}%

\subsection{Time-domain signals}
\label{sec:2.2}

To generate the total signal in the time domain from each Galactic binary population, we sum up the strain signal waveform from each binary at each time step over a one-year LISA mission time, for details see~\citet{2024arXiv240520484T}. We assume the signal's sampling frequency is 1~Hz, and the length of a signal observation from each binary is 6,311,536 observations. The sampling frequency was chosen to match the LISA sampling frequency, which is essential for resolving frequencies in the mHz range. A one-year observation time (or 365.25 days) allowed for a reasonable computation time to simulate and sample from the posterior distribution. The sum of all GW signals is our realistically simulated Galactic SGWB signal; we then add instrument noise to form the combined total LISA observations from which we will extract the SGWB signal and thus separate it from instrumental noise. We will estimate the parameters of both SGWB signal and noise ESD.

\begin{figure*}
    \centering
    \includegraphics[width=2\columnwidth]{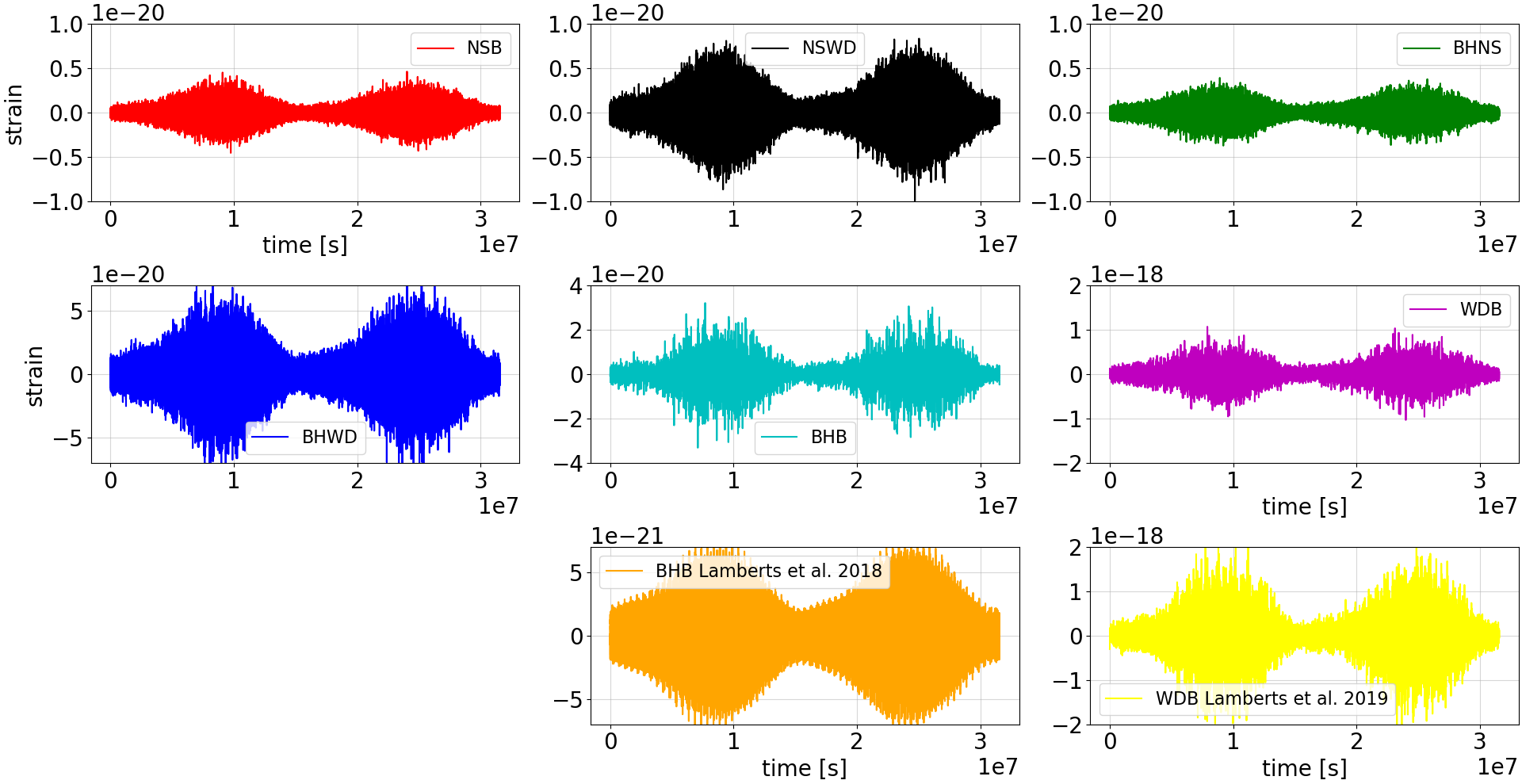}
    \caption{ Modulated signals over one-year LISA mission for a single channal A, signal frequencies ranging from 0.1~mHz to 0.1~Hz.}
    \label{fig:modulated}
\end{figure*}

The GW signal of a Galactic binary detected by LISA will be modulated over the mission time of one year due to the motion of the Earth and the LISA spacecraft around the Sun. The variation of the strain amplitude depends on the distance to the binary system and the orientation of the binary orbit with respect to the LISA detector. Combining these individual variations creates a characteristic modulation pattern due to the anisotropy of the Galaxy. For a Galactic binary, the sinusoidal modulation of the GW signal will have a period of approximately one year. The total waveform in one of the three Time-Delay Interferometry (TDI)\footnote{\label{note1}\url{https://www.cosmos.esa.int/documents/678316/1700384/SciRD.pdf}}  channels X, Y, and Z, resulting from the combination of $N$ binaries 

\begin{equation}
s_{mod}(t)= \sum_{i=1}^{N} \sum_{A=+, \times} h_{A, i}\left(t\right) \times F_{A}(\theta_i(t), \phi_i(t))\textbf{D}_A: \textbf{e}_A
\end{equation} 
where $F_A$ is the detector response function for the polarisation $A=+,\times$ to signal $i$ and $(\theta_i(t), \phi_i(t))$ describes the source location at time $t$, $h_{A,i}(t)\textbf{e}_A$ is the tensor of the amplitude of the GW , $h_{A,i}$ the dimensionless GW amplitude of the $i$th binary  and $\textbf{D}$ the one-arm detector tensor \citep[detailed in studies such as][]{2001CQGra..18.3473C, 2021MNRAS.508..803B}. This combined modulated signal is shown in Figure~\ref{fig:modulated}.  As a result, the time series of LISA observations is non-stationary. Shown in Section~\ref{sec:7SNR}, we removed sources that have SNR$>$7 and found that the amplitude of the modulated SGWBs are reduced as expected. In this study, we do not remove sources with SNR $>$ 7 as these sources are not guaranteed resolvable when they are in the same frequency bins.

As shown in our recent paper~\citet{2024arXiv240520484T}, Figure~\ref{fig:modulated} shows that the BPASS WDB population has the strongest signal, note the change in the y-axis scale between populations; the strongest signal is in the order of $ 2 \times 10^{-18}$, which is a couple of magnitude stronger than other BPASS Galactic binary populations. This trend is reasonable as WDBs are the most numerous binaries in the galaxy and are the main target for LISA. The BPASS BHWD population is the second most prominent signal amongst all the Galactic populations, even though the BHWD population is less numerous than the NSWD population; a similar trend is also found in star cluster by \citet{2023MNRAS.524.2836V}. The NSWD population also has stronger strain and more systems than the NSB, NSWD and BHB populations. 

The demodulation process can be complex and requires careful calibration and modelling of the LISA detector and the GW signal. An approach to demodulating the GW signal seen by LISA can involve removing the Doppler modulation induced by the motion of the Earth and the LISA spacecraft around the Sun by applying a time-varying phase correction as suggested in \citep{2003CQGra..20S.163C}. An alternative approach by 
\citet{2025PhRvD.111f3005P} is based on a characterisation of the cyclostationarity of the Galactic SGWB signal and  provides a purely frequency-based method to study LISA’s capability to detect the MW foreground and SGWBs from the most promising Milky Way satellites.
Another recent study by \citet{2025PhRvD.111b3025C} directly treats its anisotropy via astrophysically motivated templates, connecting the observed time modulation of the foreground amplitude and the underlying spatial distribution of the Milky Way. 

\begin{figure*}
    \centering
    \includegraphics[width=2\columnwidth]{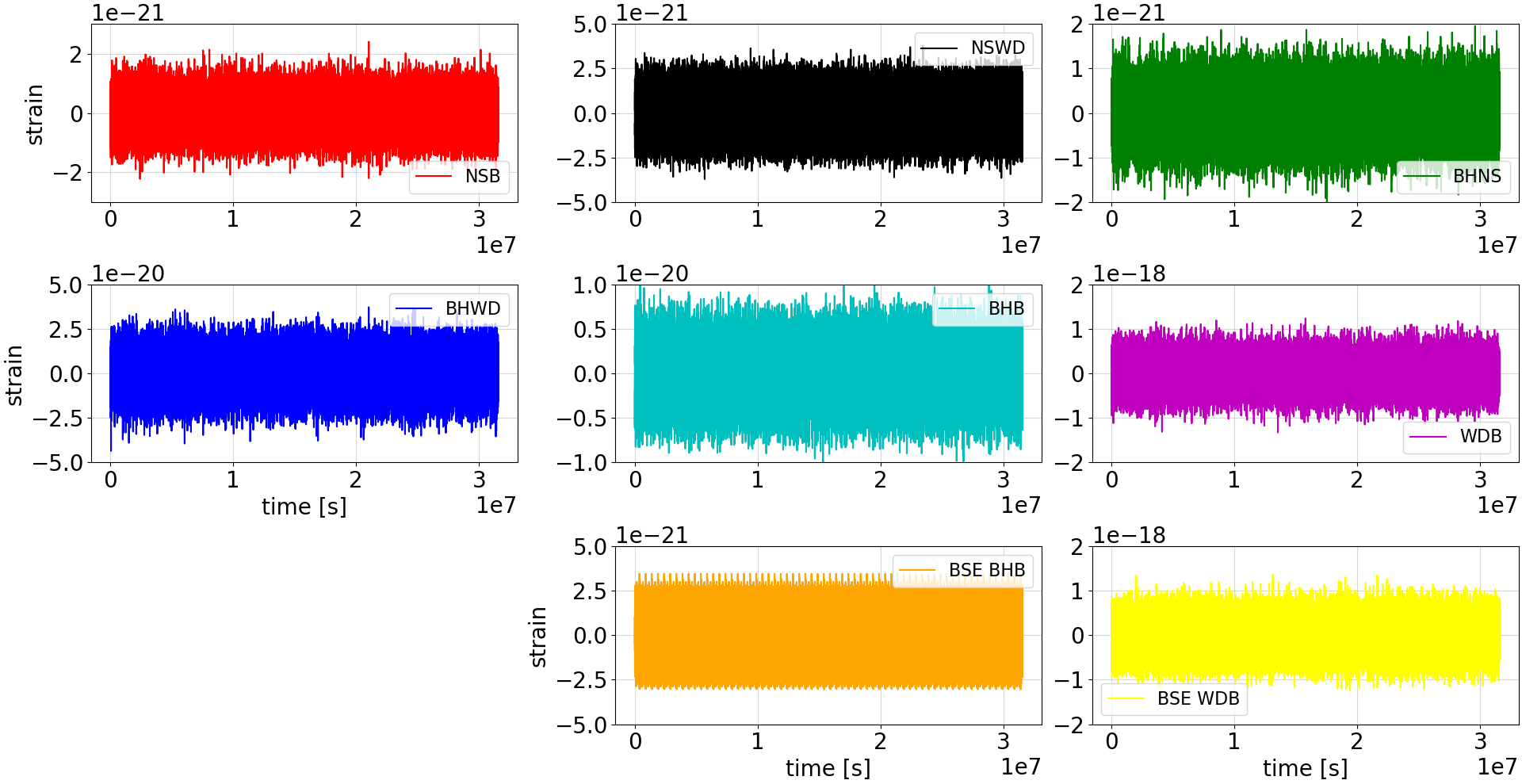}
    \caption{ Response function $\overline{F}_{A}(t)$ averaged signals over one-year LISA mission, signal frequencies ranging from 0.1~mHz to 0.1~Hz.}
    \label{fig:averaged}
\end{figure*}

In this study, for each individual signal waveform, we average the response function $\overline{F}_{A}$ over all time-varying parameters $\theta$ and $\phi$ for both polarisations, where the source location $\theta$ and $\phi$ are related to the angles of the LISA transfer function as discussed in \citet{2001CQGra..18.3473C}. As described in the paper, we linearly combine the two polarisations of the signal to get a total signal. Finally, we sum up $N$ individual signal waveforms: 
\begin{equation}
\begin{aligned}
\label{d_t}
s(t)=& \sum_{i=1}^{N} \sum_{A=+, \times} h_{A, i}(t)\textbf{D}_
A:\textbf{e}_A \times \overline{F}_{A}.
\end{aligned}
\end{equation}
The results are stationary time series strain signals shown in Figure~\ref{fig:averaged}. 

While averaging the response function in GW data analysis for LISA can simplify the statistical analysis, it is essential to note that this approach is typically only applicable when working with simulated data where each individual GW signal has been simulated. In the case of real LISA data, the situation is more complex due to that LISA only observes the total Galactic binary population signal, so one can not average the response function for each individual binary. 

To estimate the ESD of the SGWB for LISA, we first show how we generate the LISA noise and add this to the Fourier transformed GW signal in the frequency domain in Sections~\ref{sec:3.1} and \ref{sec:3.2}. We then define Bayesian models that relate the noise and GW ESD's to the simulated data in the frequency domain, and define the priors and likelihood specifications in Section~\ref{sec:4}. 
\section{Frequency-domain signals}
\label{sec:3}

A time domain observation $d(t)$ is composed of GW signal $s(t)$ plus instrumental noise $n(t)$, and we assume the observations to be a stationary time series which is normally distributed with mean 0 and covariance matrix $\Sigma$. Note that $s(t)$ is the combined signal from all binaries as in Equation~\ref{d_t}. To avoid the estimation of the high-dimensional covariance matrix $\Sigma$ in the time domain, we Fourier transform the time-domain GW signal $s(t)$ to:
\begin{equation}
\label{tilde_d_t}
\tilde{s}(f_k)=\sum_{i=1}^T s(t_i) e^{-2 \pi i t_i f_k}, 
\end{equation}
where $f_k = 2\pi k/T$ are the Fourier frequencies and $k = 0,\dots, N=T/2 -1$, and $s(t_i)$ is the time-domain signal at time point $t_i$, $i=1,\ldots,T$. For a single LISA signal channel, the total  signal $\tilde{d}(f_k)$ that LISA will observe is the sum of Fourier transformed noise $\tilde{n}(f_k)$ and suitably scaled GW signal $\tilde{s}(f_k)$:
\begin{equation}
\label{Di}
\tilde{d}(f_k) = \tilde{n}(f_k) + \tilde{s}(f_k).
\end{equation}
For the SGWB, it is common practice to predict the signal in terms of the energy spectral density per logarithmic frequency interval scaled by the critical density which relates to the power spectral density as follows:
\[h^2\Omega_{GW}(f)=h^2 \frac{4\pi^2}{2H_0^2}f^3 S_{GW}(f)
\]
where $\rm H_0$ is the Hubble-Lemaître constant ($\rm H_0 \simeq 2.175 \times 10^{−18} s^{-1}$). 
Here, the frequency domain data in (\ref{Di}) are scaled in units of the energy density.

In Section~\ref{sec:3.1}, we explain in detail how to generate LISA noise in the frequency domain. Due to the normalising and decorrelating effect of the Fourier transform, the Fourier coefficients $\tilde{d}(f_k)$ are asymptotically independent and complex Gaussian with mean zero and variance given by the sum of the ESDs of the signal and noise. This is the basis of the Whittle likelihood approximation, the product of independent complex Gaussian variables with mean 0 and variance equal to the ESD $h^2\Omega_{tot}(f_i)$, quantifying the probability of observing the data given the noise and GW signal ESDs:
\begin{equation}
\label{likelihood}
\mathcal{L}(\tilde{d}) \mid 0, h^2 \Omega_{\rm tot}(f_i)) \propto  \prod_{i=1}^{N} \frac{1}{h^2 \Omega_{\rm tot}(f_i)} \exp\left(-\frac{|\tilde{d}(f_i)|^2|}{h^2\Omega_{\rm tot}(f_i)}\right),
\end{equation}
where $h^2\Omega_{\rm tot}(f_i)$ represents the sum of $h^2\Omega_n(f_i)$ and $h^2\Omega_{GW}(f_i)$: 

\begin{equation}
\label{h_total}
h^2 \Omega_{tot}(f_i) = h^2 \Omega_n(f_i) + h^2 \Omega_{GW}(f_i).
\end{equation}

We use different power law functional forms to model the ESD of the GW signal $h^2 \Omega_{GW}(f)$. This will be discussed in Section~\ref{sec:3.2}. Firstly, we outline how we simulate LISA noise. 

\subsection{LISA noise}
\label{sec:3.1}

We adopt the noise model specified in ESA's science requirement document\footnote{\label{note1}\url{https://www.cosmos.esa.int/documents/678316/1700384/SciRD.pdf}}, Derivations and details can be found in \citet{2019CQGra..36j5011R} and \citet{2019PhRvD.100j4055S}. This noise model is based on the results obtained from the successful LISA Pathfinder mission, which demonstrated effective noise control in the frequency range of approximately $3\times 10^{-5}$ to 1~mHz \citep{2018PhRvL.120f1101A}. In our study, we extend the frequency range up to 0.1~Hz, which allows us to study binary systems at higher frequencies for LISA. We assume the noise ESD is a given functional form.

To simulate LISA instrument noise, we take into account noises related to LISA lasers and optics \citep{2019JCAP...11..017C}., where the noise level of a single TDI channel T can be determined within an uncertainty of around 20 per cent of the true noise parameter values across the frequency range of $3\times 10^{-5}$ to 0.5~Hz \citep[][]{2019JCAP...11..017C}. We note that in another study by \citet{2024PhRvD.109d2001M}, the authors point to remaining uncertainty about the instrumental noise model, which we will not consider in our work for simplicity.

Our assumptions on the LISA instrument noise include constant and equal arm lengths, with equal noise power spectral densities (PSD). Consequently, all noise components are organised into two effective functions \citep{2019JCAP...11..017C}. The high-frequency noise components, which dominate at higher frequencies, are represented by the one-link "optical metrology system" noise PSD $P_{\mathrm{oms}}(f, P)$:
\begin{equation}
\label{P}
P_{\rm oms}(f, P)=P^2\left[1+\left(\frac{2}{f}\right)^4\right]\left(\frac{2 \pi f}{c}\right)^2
\end{equation}
where c is the speed of light.
The low-frequency components are accounted for by the single "mass acceleration" noise PSD $\rm P_{acc}(f, A)$:
\begin{equation}
\label{A}
P_{\rm acc}(f, A) =A^2 \left[1+\left(\frac{0.0004}{f}\right)^2\right]\left[1+\left(\frac{f}{0.008}\right)^4\right]\left(\frac{1}{2 \pi fc}\right)^2,
\end{equation}
where the unit of $P$ is in $pm$ and $A$ is in $fm$. The simulated noise is generated using $P$ = 15 and $A$ = 3, and we treat them as unknown parameters to be estimated. To combine the two noises, the single TDI X channel  PSD $P^X_n(f, P, A)$ is defined as:
\begin{equation} 
P_n^X(f, P, A)=16 \sin ^2\left(\frac{2 \pi f L}{c}\right)\left\{P_{\rm oms}+\left[3+\cos \left(\frac{4 \pi f L}{c}\right)\right] P_{\rm acc}\right\},
\end{equation}
where L is the constant LISA arm length of 2.5 million km. To obtain the noise PSD, we consider $\mathcal{R}(f)$, the sky-averaged detector response function \citep[][]{2018CQGra..35p3001C} approximation for the $X$, $Y$ and $Z$  channels in the frequency domain, which is defined as:
\begin{equation}
\mathcal{R}(f) \simeq 16 \sin ^2\left(\frac{2 \pi f L}{c}\right) \frac{3}{10} \frac{1}{1+0.6(2 \pi f L / c)^2}\left(\frac{2 \pi f L}{c}\right)^2.
\end{equation}
Then the noise PSD is calculated with the response function $\mathcal{R}(f)$ as:
\begin{equation}
S_n(f)=\frac{P_n}{\mathcal{R}(f)}.
\end{equation}

Finally, we can express the instrument noise in terms of the noise ESD $h^2\Omega_n(f)$ \citep[][]{2018CQGra..35p3001C} as:
\begin{equation}
h^2\Omega_n(f)=h^2\frac{4 \pi^2}{3 H_0^2} f^3 S_n(f).
\end{equation}

Now that we have the noise ESD $h^2\Omega_n(f)$, we can generate the real and imaginary parts of the LISA instrument noise at each frequency. The noise is assumed to follow a complex Gaussian distribution with a mean of 0 and a variance equal to the noise ESD at each frequency. The method is outlined in \citet{2025PhRvD.111b3025C}. Since the GW signal and the noise are uncorrelated, we generate the noise components directly and incorporate them into the total data stream, $\tilde{d}(f_i)$, the sum of the GW signal and noise. 

\subsection{LISA Galactic binary GW background}
\label{sec:3.2}

\begin{figure*}
    \centering
    \includegraphics[width=2\columnwidth]{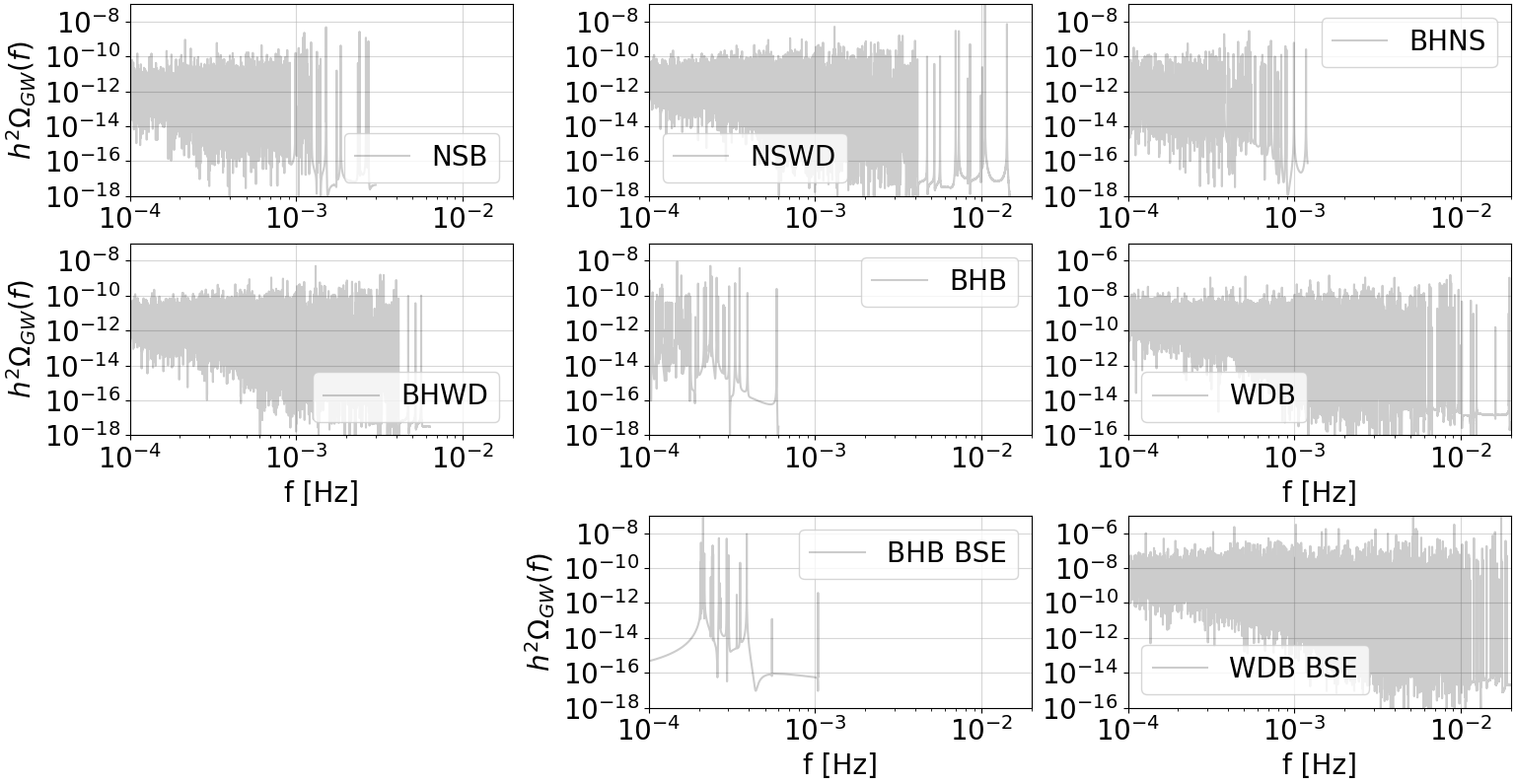}
    \caption{The periodogram of the GW signal of different Galactic binary populations.}
    \label{fig:ESD}
\end{figure*}

The periodogram  $I_s(f_k)=|\tilde{s}(f_k)|^2$ is defined as the squared modulus of the discrete Fourier transformation of the time-domain signal. Figure~\ref{fig:ESD} shows the periodogram of the combined GW signals from different Galactic populations. There is a cutoff frequency 30~mHz near the high end of the LISA frequency range for the BHNS and BHB populations, as after these cutoff points, the BHNS and BHB systems may have either already merged or moved beyond the LISA frequency region, i.e., into the LIGO-Virgo-KAGRA (LVK) kHz frequency range. BPASS NSB, BHNS, and BHB populations have power only  at frequencies below 2~mHz \citep[][]{2020LRR....23....3A}. On the other hand, BPASS NSWD, BHWD and WDB populations have signals extended above 1~mHz, where LISA is most sensitive. Besides the WDBs systems, BHNSs  in the range of 2~mHz to 10~mHz also contribute to the guaranteed astrophysical components in the LISA band \citep[discussed in][]{2019PhRvD.100f1101A}. However, our BHNS population does not have systems at these frequencies. We expect this to result from our BH mass distribution extending to higher masses, which implies the BHNS have shorter merger times and, thus, a dearth of binaries with short periods.

Furthermore, shown in Figure~\ref{fig:ESD}, compared with the BHB from the study by \citet{2018MNRAS.480.2704L}, the BPASS BHBs have more power at the lower end of the frequency range, i.e. $\leq 0.2$~mHz. These BPASS BHB systems have a bigger orbital period than systems with higher frequencies. As for the BPASS WDB population, compared with the BSE WDB \citep[][]{2019MNRAS.490.5888L}, we see similar shapes, but BPASS WDBs produce weaker signals and fewer signals above 10~mHz. We note that the `popcorn-like' features \citep[][]{2023PhRvD.107j3026L} of the BHB systems, where the power of signals exhibits visible spike-like feature, suggest that individual GW from BHB sources may be identifiable. However, identifying these individual GWs would require a different technique, which is not the focus of this work.

We specify three different functional forms for the SGWB ESDs shown in Eqs.~\ref{PL}, \ref{Broken}, and \ref{Peak} and estimate their parameters. We stress that these GW ESD functional forms only approximations the true GW ESDs. In the following, we detail the three different parametric models for SGWB and noise ESD.

Many studies \citep[e.g.][]{2019JCAP...11..017C,2021MNRAS.508..803B} have used the simplified assumption that the SGWB ESD can be effectively described by a single power law (PL) in the frequency domain, which is defined by its amplitude and slope. Thus, based on references \citet{2013PhRvD..88l4032T, 2015MNRAS.453.2576L, 2018ApJ...859...47A, 2019PhRvD.100f1101A}, we define the ESD of the SGWB as a single PL form with two parameters, namely amplitude $\alpha$ and slope $\beta$ as:
\begin{equation}
\label{PL}
h^2\Omega_{\rm GW}^{\rm PL}(f) = \alpha\left(\frac{f}{f_{\rm ref}}\right)^\beta,
\end{equation}
where $f_{\rm ref}$ is the reference frequency of 0.001~Hz. Combined with the noise parameters $P$ and $A$ in Eq.~\ref{P} and \ref{A}, overall, this model depends on four parameters ($P$, $A$, $\alpha$ and $\beta$). 

A single PL is limited to describing the GW signals \citep{2018JCAP...11..038K, 2018PhDT.......213M}, as the stochastic nature of the signals. In the hope of capturing both low and high-frequency ESD shapes, we define a broken PL shape for the GW signal while keeping the same noise models. This distinctive spectral pattern may originate from the amalgamation of two distinct physical sources, as discussed in \cite{2019JCAP...11..017C}. The broken PL $ h^2\Omega_{\rm GW}^{\rm Broken}(f)$ ESD is thus defined as:
\begin{align}
\label{Broken}
h^2\Omega_{\rm GW}^{\rm Broken}(f) = & \, \alpha \left[ \left( \frac{f}{0.002} \right)^\gamma \Theta(f_T - f) \right. \nonumber \\
& \left. + \left( \frac{f_T}{0.002} \right)^\gamma \left( \frac{f}{f_T} \right)^\delta \Theta(f - f_T) \right]
\end{align}
where  the constant $f_T$ denotes the change-point of the slope and $\Theta$ is the Heaviside step function. We note that the second model depends on six parameters $P$, $A$, $\alpha$, $\gamma$, $\delta$ and $f_T$.

In addition to the single PL and the broken PL, phenomena like the Cosmic Microwave Background (CMB) \citep[such as]{2016JCAP...01..041N} and primordial BHs \citep[][]{2024PhRvD.109d2001M} generate SGWB signals that exhibit a singular peak within the frequency range detectable by LISA. Sources originating during inflation, such as those discussed in \cite{2016JCAP...01..041N}, also generate SGWB signals exhibiting pronounced amplification at distinct frequencies within the LISA bandwidth. These signals can be accurately characterised using a functional form resembling a single-peak model. A recent study by \citet{2023MNRAS.524.2836V} based on the BPASS population for globular clusters indicate that each binary population has a maximum strain, with most maximum powers occurring within the LISA frequency band. Their study may indicate that a broken PL or a single-peak spectral shape is more appropriate for describing the Galactic binary population ESDs. As in \cite{2019JCAP...11..017C}, we define the single-peak GW ESD functional form as:
\begin{equation}
\label{Peak}
h^2\Omega_{\rm GW}^{\rm Peak}(f)=\alpha \operatorname{Exp}\left\{-\frac{\left[\log _{10}\left(f / f_b\right)\right]^2}{\Delta^2}\right\},
\end{equation}
where $f_b$ is the central frequency and $\Delta$ is the width of the curve. Our Bayesian model contains five parameters: $\alpha$, $f_b$, $\Delta$ and the two noise parameters $P$ and $A$. 

The Bayesian model is completed with the prior distributions on the unknown parameters, which will be specified in Section~\ref{sec:4}.

\section{Bayesian inference}
\label{sec:4}
Bayesian statistics are widely used in and well suited to GW astronomy. In the Bayesian framework, we combine prior knowledge  with observed data to make probabilistic inferences about physical phenomena. Our Bayesian estimation pipeline relies on Markov chain Monte Carlo (MCMC) techniques for computing the posterior distribution. For a detailed summary of parameter estimation for GWs signals, see \citet{2013PhRvD..88l4032T, 2022RvMP...94b5001C}. 

For estimating the GW parameters, we treat the GW signal ESD and noise parameters, combined into the parameter vector $\theta$, as random variables. The parameter vector $\theta$ depends on the ESD model we use (e.g. a single PL $\theta$ is $A, P, \alpha$, and $\beta$). All knowledge about the unknown parameters is used to specify the prior distribution $p(\theta)$. Bayes theorem tells us that given the observed data $d$ -- in our case, $d$ is the LISA signal plus noise that we simulated -- we can update the prior distribution through the likelihood, $p(d|\theta)$, to the posterior distributions $p(\theta|d)$ of the parameters by
\begin{equation}
\label{eq:bayes}
p(\theta|d)=\frac{p(d|\theta) p(\theta)}{p(d)}.
\end{equation}

On the right-hand side of this equation, we multiply the likelihood $p(d|\theta)$ (defined in Eq.~\ref{likelihood}) with the prior distribution $p(\theta)$ of the parameters. The term $p(d)$ is the evidence or marginal likelihood that does not depend on the parameters $\theta$ and can be seen as a normalisation constant that does not need an explicit calculation; however, it would be necessary for model comparison. Even though the analytical  form of the posterior $p(\theta|d)$ is known, the difficulty is to marginalise this multivariate distribution of the parameter vector $\theta$ to obtain  the  posterior distributions of each individual parameter. Therefore, numerical sampling algorithms are often used to obtain samples $\theta^{(i)} \sim p(\theta|d), i=1,\ldots,N$ for some large number $N$. The sampled values of each individual parameter can then be used to estimate characteristics of its marginal posterior distribution.
Typically, computing the posterior distribution is done  based on Markov chain Monte Carlo algorithms.

To validate our Bayesian approach, we performed simulation studies shown in Appendix~\ref{sec:simulation}. In the simulation studies, we generated noise and GW signals using parametric ESD models and showed that the models performed well in estimating the ESD parameters. 
 
The priors for the noise parameters $P$ and $A$ are set to be Gaussian, centered at 15 and 3 with standard deviations 3 and 0.6, respectively, which means that we assume that the noise parameters are known within $\pm$ 20 per cent. This corresponds to the prior assumptions in \citet{2018CQGra..35p3001C} and in the simulation study in Table~\ref{tab:1}. 

As the GW strain varies considerably in the LISA frequency band, we split the frequency range into two sections: a lower frequency range from 0.1~mHz to 1~mHz and a higher frequency range from 1~mHz to 30~mHz, respectively. We choose 1~mHz as the split point because no BHs are present above 1~mHz in our Galactic populations, shown in Figure~\ref{fig:frequency}. These systems with BHs have likely already merged or moved to a higher frequency band. The same Figure shows the WD dominant GW signal at frequencies above 1~mHz. We also estimate the ESD over the full LISA frequency band for comparison.

Prior distributions for all the populations are shown in Table~\ref{tab:3}. Regardless of whether we estimate the ESD in the low or high frequencies, the priors are the same for both frequency ranges. 

\renewcommand{\arraystretch}{1.5}
\begin{table}
\fontsize{7pt}{7pt}\selectfont
  \centering
  \caption{ Priors for single PL (top), broken PL (middle) and single-peak (bottom) models that are used in our BPASS Galactic binary populations Bayesian models.}
  \begin{tabular}{|c|c|c|c}
     & \multicolumn{1}{c}{\textbf{\underline{Single PL}}} \\
        $\alpha$ & $\beta$  \\
        U($10^{-16}$, $10^{-6}$) & Norm(0, 1) & \\
     \hline        
     & \multicolumn{1}{c}{\textbf{\underline{Broken PL}} }\\
        $\alpha$ & $\delta$ & $\gamma$ & $f_T$ \\
        HalfNorm($10^{-11}$) & Norm(0, 1) & Norm(0, 1) & U($10^{-4}$, $3\times 10^{-2}$)\\
    \hline        
     & \multicolumn{1}{c}{\textbf{\underline{Single-peak}} }\\
        $\alpha$ & $\Delta$ & $f_b$ \\
        HalfNorm($10^{-11}$) & U(0, 15) & U($10^{-4}$, $3\times 10^{-2}$)\\
    \label{tab:3}
  \end{tabular}
\end{table}

For the amplitude $\alpha$, the HalfNormal distribution was used to avoid negative values, parametrised by its standard deviation. The Uniform distributions are relatively uninformative and cover a wide range of frequencies, reflecting a large prior uncertainty. 

We note that in the broken PL model, we let the switch frequency $f_T$ and curvature of the single-peak $f_b$ vary rather than fixing their values \citep[as was done in the studies by][and our test model in Table~\ref{tab:1} as examples]{2018CQGra..35p3001C, 2021MNRAS.508..803B}. This is because $f_T$ and $f_b$ are likely to be different for different Galactic binary populations; this claim is based on a study by \citet{2023MNRAS.524.2836V} where it was found that different binary population GW signals peak at different frequencies in globular clusters. In addition, another study by \citet{2022ApJ...924..102R} also suggests that there is a maximum in BHB signal strain in the frequency domain.

Many tools can be used for a Bayesian computations; the procedure explained in this section is implemented using the Python library PyMC3 \citep[][]{2015arXiv150708050S, 2016arXiv160700379C}. For detail of the Bayesian analysis tools see Appendix~\ref{sec:4.2}.
\section{Results}
\label{sec:5}

This section presents estimated Galactic binary population ESDs, assuming a single PL model. For interested readers, the parameter estimates for all models are provided in Table~\ref{tab:model_comparison}. 

In all our models, the noise parameters $P$ and $A$ have been estimated with reasonable accuracy compared to the known parameter values of $P=15$ and $A=3$. These estimates are consistent across all models and populations, as shown in Figures~\ref{fig:pl_test1}, \ref{fig:broken_test_cornor} and \ref{fig:peak_test_cornor}. We also note that the differences between the estimates $P$ and $A$ of the different models are minimal. Therefore, we only present results for a single PL ESD of the GW signal without including plots of the noise ESD.

\subsection{Individual Galactic population results}
\label{5.1}

\begin{figure*}
    \centering
    \includegraphics[width=\textwidth]{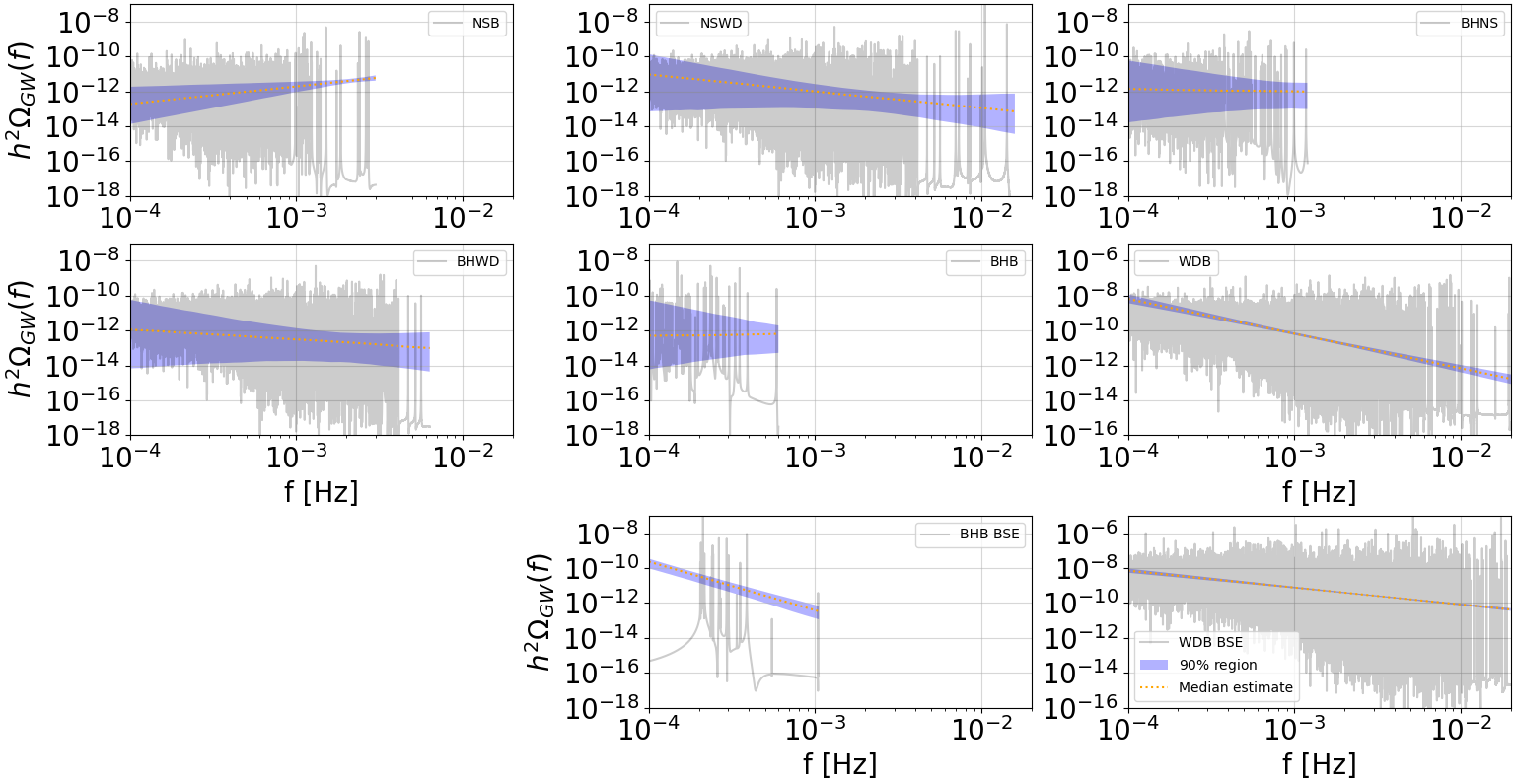}
    \caption{ Four-parameter Bayesian parametric single PL models for estimating ESD over the full LISA frequency range of 0.1~mHz to 30~mHz. The grey lines represent the periodogram of the GW signal. The orange lines represent the median ESD estimates, and the HDI is taken as 90 per cent is shaded in blue.}
    \label{fig:estimates}
\end{figure*}

Our results show how the ESD estimates differ between different Galactic populations. In Figure~\ref{fig:estimates}, we show the median PL ESD estimates accompanied with the highest density interval (HDI) taken as 90 per cent (in blue) over the LISA frequency range 0.1~mHz to 30~mHz. The 90 per cent region for the BPASS WDB population is narrower than the other populations because the WDB population has the strongest signals and more observations. Therefore, the uncertainty of the estimate is smaller than that of other populations.

The posterior means for all the parameters in the single PL Bayesian model and values of $h^2 \Omega_{\rm GW}$ at 3~$\text{mHz}$ are displayed in Table~\ref{tab:combined}. Notably, our prediction of Galactic BHB population $h^2 \Omega_{\rm BHB} = 2.9 ^{+17}_{-2.2} \times 10^{-12}$ is similar to the prediction by \citet{2019ApJ...871...97C} of $\Omega_{\rm BHB}=2.7^{+2.8}_{-1.6} \times 10^{-12}$; and our prediction of Galactic NSB population $h^2 \Omega_{\rm NSB}=1.2 ^{+3.5}_{-4.0} \times 10^{-12}$ is also comparable to the prediction by \citet{2019ApJ...871...97C} of $\Omega_{\rm NSB}=1.7^{+3.5}_{-1.4} \times 10^{-12}$. Our Galactic WDB population ESD estimate $h^2 \Omega_{\rm WDB} = 18 ^{+2}_{-2} \times 10^{-12}$ at 3~\text{mHz} is in the range of that predicted in LMC\footnote{We note that the LMC value may not be useful, as it is for a completely different galaxy than the Milky Way, but we mention the values for a reference purposes.} $\Omega_{\rm WDB} \approx 10^{-11}$ at 1~\text{mHz} suggested by \citet{2023arXiv230812437R} and $\Omega_{\rm WDB} = 4 \times 10^{-11}$ at 1~\text{mHz} suggested by a recent study on the astrophysical gravitational wave background (AGWB) for LISA by \citet{2024arXiv240710642H}. 

We also offer our predictions for the $h^2 \Omega$ of the BHNS, BHWD and NSWD populations in Table~\ref{tab:combined}. All three population $h^2 \Omega$ values are on the order of $10^{-12}$. Overall, we predict the WDB and BHB populations have the strongest $h^2 \Omega$ at 3~$\text{mHz}$. All the other populations have similar $h^2 \Omega$ at 3~$\text{mHz}$, which may mean that it is difficult to tell these populations apart as the GW signals coming from them will overlap.

While these comparisons provide useful context, it is important to acknowledge that the referenced studies primarily consider extragalactic foregrounds, whereas our study focuses on Galactic sources. This distinction may influence the direct comparability of the results. However, the similarities in the predicted values suggest that our Galactic estimates are within a reasonable range of those derived from extragalactic studies, providing additional validation for our model. It is also important to note that our Galactic prediction comparisons are limited by the lack of observational data and the binary population synthesis models used. These limitations should be considered when interpreting the results, as they may affect the precision and applicability of our estimates.

\begin{table*}
  \centering
  \caption{Posterior distributions for all the parameters in the single PL Bayesian model, for broken PL and single-peak (or Gaussian bump) model estimates see Table~\ref{tab:6}. The mean estimates are accompanied with the HDI is taken as 90 per cent. References 1-3: \citet{2021MNRAS.508..803B,2023arXiv230812437R, 2024arXiv240710642H}, 4-8: \citet{2016PhRvL.116w1102S, 2016PhRvD..94j3011D, 2020MNRAS.493L...1C, 2021PhRvD.103d3002P, 2024PhRvD.109d2001M}, 9 and 10: \citet{2019ApJ...871...97C,2021PhRvD.103d3002P} }
  \begin{tabular}{ccccc}
     \hline
     Population & $h^2\Omega_{\rm GW}$ (Reference) & Amplitude $\alpha$ (Prediction) & Slope $\beta$ (Prediction)& $h^2\Omega_{\rm GW}$ (Prediction) \\
     \hline
     WDB & $h^2\Omega_{\rm WDB} \approx 10^{-9}$ (1) & $100^{+2}_{-2} \times 10^{-12}$ & $-1.56 ^{+0.03}_{-0.03}$ & $18 ^{+1}_{-1} \times 10^{-12}$ \\
          &  $h^2\Omega_{\rm WDB} \approx 10^{-11}$ (2)  &               &                   & \\
          &  $h^2\Omega_{\rm WDB} \approx 10^{-12}$ (3)  &               &                   & \\
     BHB & $h^2\Omega_{\rm BHB} \in [10^{-13}, 10^{-11}]$ (4-8) & $1.6^{+1.5}_{-0.2} \times 10^{-12}$ & $0.42 ^{+1.0}_{-1.0}$ & $2.9 ^{+1.0}_{-1.0} \times 10^{-12}$ \\
     NSB & $h^2\Omega_{\rm NSB} \approx 10^{-12}$ (9)& $1.2^{+0.6}_{-0.1} \times 10^{-12}$ & $0.00 ^{+0.40}_{-0.32}$ & $1.2 ^{+1.5}_{-3.0} \times 10^{-12}$ \\
         & $h^2\Omega_{\rm NSB} < 10^{-13}$ (10)   &               &                   &  \\
     BHNS & ---  & $1.3^{+0.2}_{-0.2}\times 10^{-12}$         & $-0.42 ^{+0.43}_{-0.44}$   & $0.9 ^{+0.6}_{-0.3} \times 10^{-12}$\\
     BHWD &  --- & $1.4^{+0.6}_{-0.3}\times 10^{-12}$         & $-0.05 ^{+0.50}_{-0.40}$   & $1.3 ^{+3.1}_{-0.4} \times 10^{-12}$ \\
     NSWD & ---  & $1.7^{+1.0}_{-1.0}\times 10^{-12}$         & $-2.38 ^{+0.06}_{-0.06}$   & $1.5 ^{+1.9}_{-1.9} \times 10^{-12}$ \\
     \hline
  \end{tabular}
  \label{tab:combined}
\end{table*}

\begin{figure}
    \centering
    \includegraphics[width=0.45\textwidth]{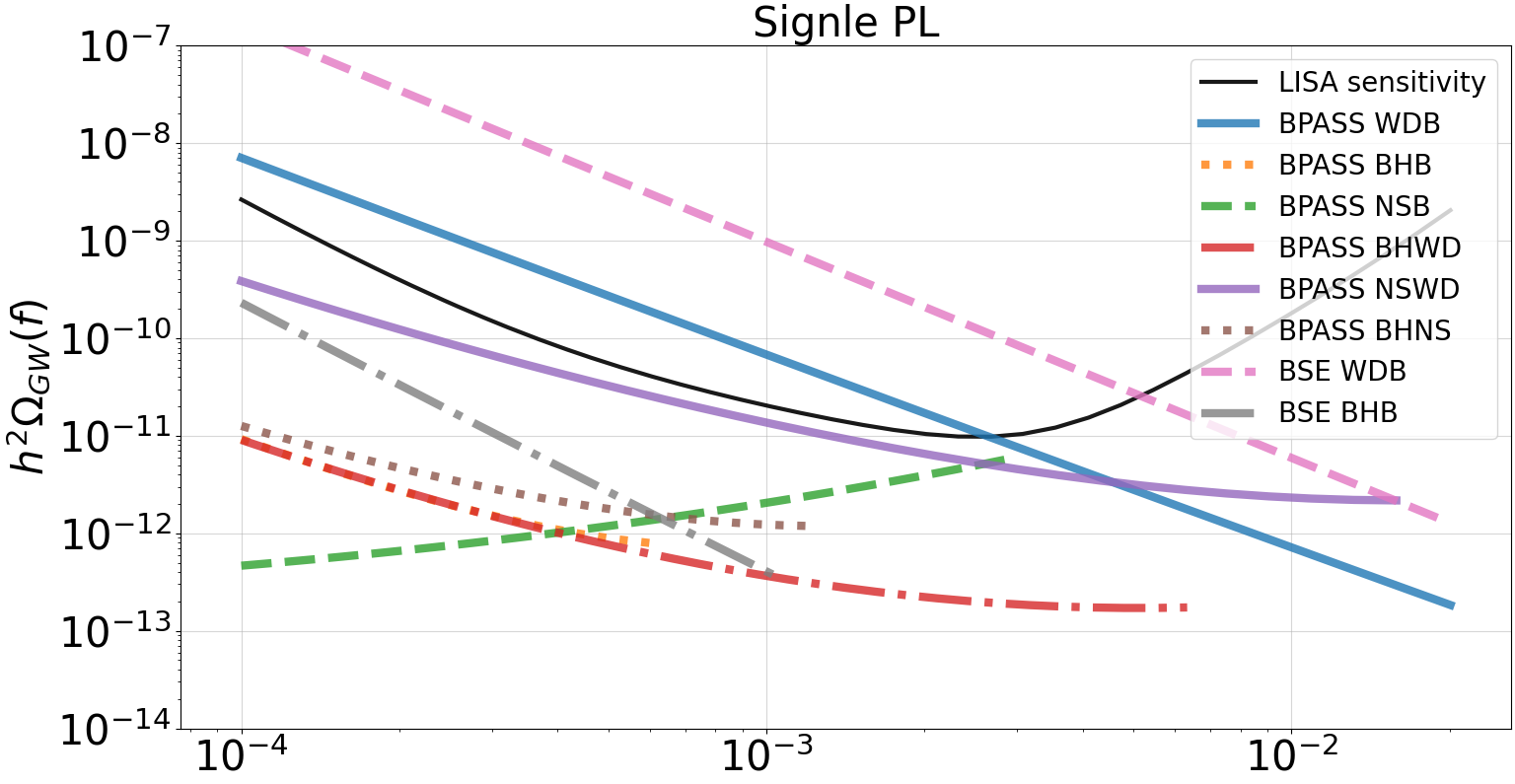} 
    \includegraphics[width=0.45\textwidth]{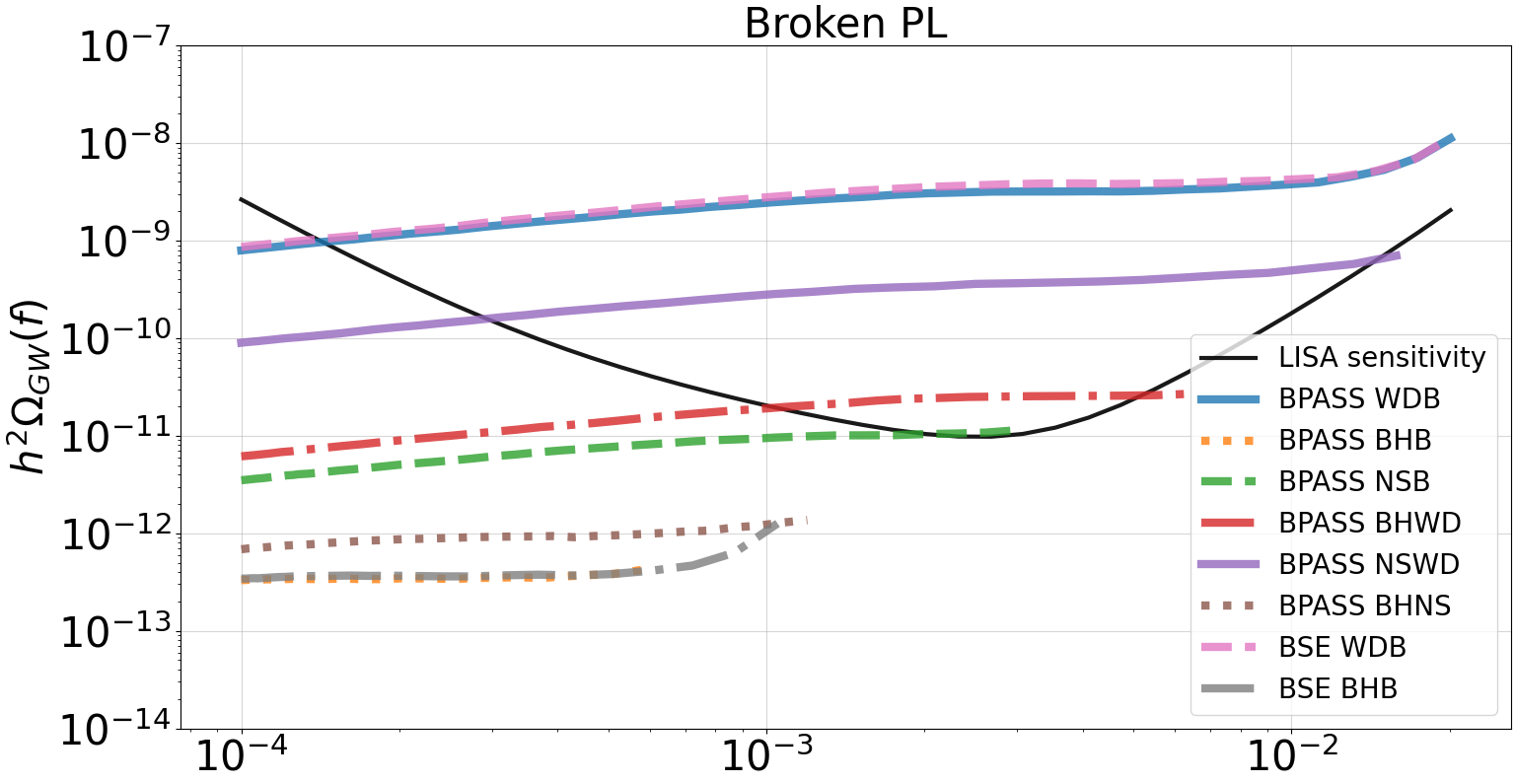}
    \includegraphics[width=0.45\textwidth]{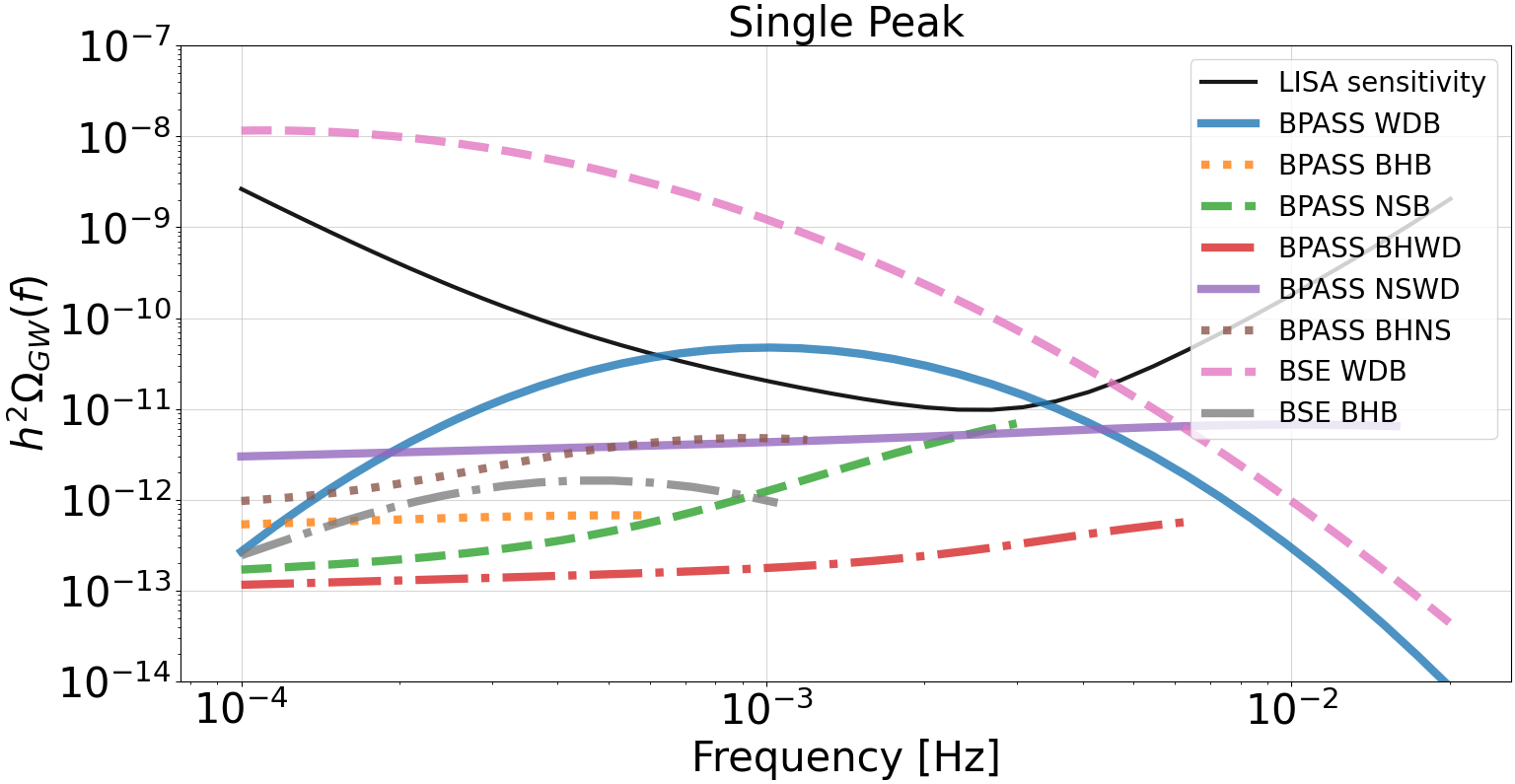}
    \caption{ Estimated ESD from mean parameter values for all Galactic populations, black line is the LISA sensitivity evaluated as $h^2\Omega_n(f)$. Top: assuming the GW signal ESD is in the form of a single PL. Middle: broken PL model. Bottom: single-peak model.}
    \label{fig:multi}
\end{figure}

In addition to the result of the single PL model, Figure~\ref{fig:multi} shows the estimated ESD from mean parameter values for all Galactic populations while assuming that the GW signal spectral shape is a single PL, broken PL and a single-peak over the LISA frequency range of 0.1~mHz to 30~mHz. As mentioned earlier, based on the GW spectral synthesis result from the recent study by \cite{2023MNRAS.524.2836V} from the BPASS population for globular clusters, the spectral shape of the Galactic populations is not likely to be a single PL. Their study suggests that each binary population has a maximum strain, and most maximum powers have frequencies within the LISA band. Therefore, a broken PL or a single-peak spectral shape is more reasonable for describing the Galactic binary population ESDs. 

To compare the different parametric models, we compute the model comparison criterion ELPD-WAIC \citep{2015arXiv150704544V} and corresponding SE values in Table~\ref{tab:model_comparison}. Smaller  ELPD-WAIC values indicate a better model fit. The underlined ELPD-WAIC values are the smallest of the three models in each Galactic binary population. Notice that no model stands out as the best model based on the ELPD-WAIC and their corresponding SE values. 
\begin{table*}
  \small
  \centering
  \caption{ Model comparison ELPD-WAIC and SE Values for Single PL, broken PL, and single peak for the Galactic populations}
    \begin{tabular}{r|rrrrrrrrr}
   Models&  WDB     &  BHB     &  NSB    &  BHWD   &  NSWD    &  BHNS  & BSE WDB & BSE BHB & BPASS total\\
    \hline
Single PL&24440.31 &62665.20&31341.88 &78044.02 &66401.75&14413.85&222675.22&61789.63 &246926.29 \\  
    SE   &74.03    &71.72   & 54.71   &130.52    &68.41     &37.22   &263.65    &70.61  & 190.50 \\ 
    \hline
Broken PL&24282.85&62665.85&31342.66&\underline{77951.59}&66402.34&14414.49&\underline{222652.25} &61789.96&\underline{246829.14}\\ 
    SE  &137.82    &71.78    & 54.96  &170.53    &68.52     &37.76   &265.20   &70.83 & 202.03 \\ 
    \hline
Single-peak&\underline{24431.8}&\underline{62665.00}&\underline{31341.83}&78053.82&\underline{66401.18}&\underline{14413.33}&222683.07&\underline{61789.25}&246924.22 \\ 
    SE  &74.72    &71.65    & 54.65  &126.56    &68.65     &37.29   &263.22   &70.67 & 191.57 \\ 

\hline        
  & \multicolumn{8}{c}{\underline{frequency $\leq f^{-3}$~Hz}} \\
Single PL&29086.41&--&58290.62&27084.47&58153.42&58226.78&28546.84&-- &27004.25 \\  
    SE   &48.39    &--   &68.03&197.85    & 160.09 &78.65  &67.46   &--  & 220.41 \\ 
    \hline
Broken PL&29086.18&--&58290.30&\underline{26633.05}&58152.59&58227.10&\underline{28512.10}&--&\underline{26518.41}\\ 
    SE  &48.97    &--    & 68.41  &452.59    & 167.66 & 79.00 &82.34   &-- & 511.59 \\ 
    \hline
Single-peak&29086.45&--&58290.41&27077.42&58154.34&58226.69 &28543.79&--&26993.81\\ 
    SE  &48.37    &--    &68.07 &195.92    & 159.81 & 78.79  &67.92   &-- & 222.23 \\ 
\hline        
  & \multicolumn{8}{c}{\underline{frequency $>f^{-3}$~Hz}} \\
Single PL&118596.76&--&14062.73 &72414.71&60504.25&138775.24 &216860.36&-- &117836.51 \\  
    SE   &122.92    &--   & 28.86   &126.92    &65.99    &  85.15 &88.19    &--  & 174.00 \\ 
    \hline
Broken PL&118597.47&--&14063.14&\underline{72404.32}&60504.33&138775.11&\underline{216847.86} &--&\underline{117804.92}\\ 
    SE  &123.03    &--    & 30.01  &141.09    &66.34     & 85.55  &223.55   &-- & 203.42 \\ 
    \hline
Single-peak&\underline{118596.50}&--&14062.76&72411.92&\underline{60503.32}&138775.17 &216849.28&--&117834.62\\ 
    SE  &122.93    &--    & 29.08  &127.83    &66.09    & 85.13  &222.61   &-- & 177.02 \\    
    \end{tabular}%
  \label{tab:model_comparison}%
\end{table*}%

Nonetheless, we see discrepancies between the ESD in all Galactic populations in Figure~\ref{fig:multi}. In the top figure of the single PL model, the mean ESD of WDB, NSWD and NSB populations are above or near the LISA sensitivity curve, which may indicate that some of these systems are loud enough and can be detected by LISA and the signals seem to decrease in strength towards higher frequency. In the middle figure of the broken PL model, the BHB and BHNS systems are below the LISA sensitivity curve, and their signals may remain in the foreground noise; the signal strength increases slightly towards higher frequency. In the bottom figure of the single-peak model, WDB, BHWD, and NSB populations produce strong signals above or near the LISA sensitivity curve. As we can see, the mean ESD estimate of all the populations varies considerably; the only consistency is that of the WDB population mean, which is above the LISA sensitivity curve in all models; the single-peak model results show mixed similarities of the other two models apart from the BPASS WDB population which peaks at around 1~mHz.

Moreover, in the same Figure~\ref{fig:multi}, we notice the BHB (in orange) posterior mean of the broken PL ESD confirms our recent work \citep[][]{2024arXiv240520484T}, which shows that there will be a few BHB systems above the LISA sensitivity. However, the single PL and the single-peak mean estimates indicate that the BHB population may be too quiet for LISA. The WDB (in blue) population's broken PL estimates align with our prediction, as demonstrated using \textsc{PhenomA}\footnote{A Python package for simulating LISA observations \citep[]{2007CQGra..24S.689A, 2019CQGra..36j5011R}} in our recent paper \citep[][]{2024arXiv240520484T}, for the WDB systems above the LISA sensitivity curve. Notably, the NSB (in green) population peaks at  LISA's most sensitive frequency of 3~mHz. Surprisingly, only the broken PL model shows that the BHWD population is above the LISA sensitivity, even though the BHWD population has the second strongest signal. We also noticed that all three ESD estimates of the Galactic BHB, BHWD, BHNS and NSWD populations are relatively flat.  

\begin{figure*}
    \centering
    \includegraphics[width=2\columnwidth]{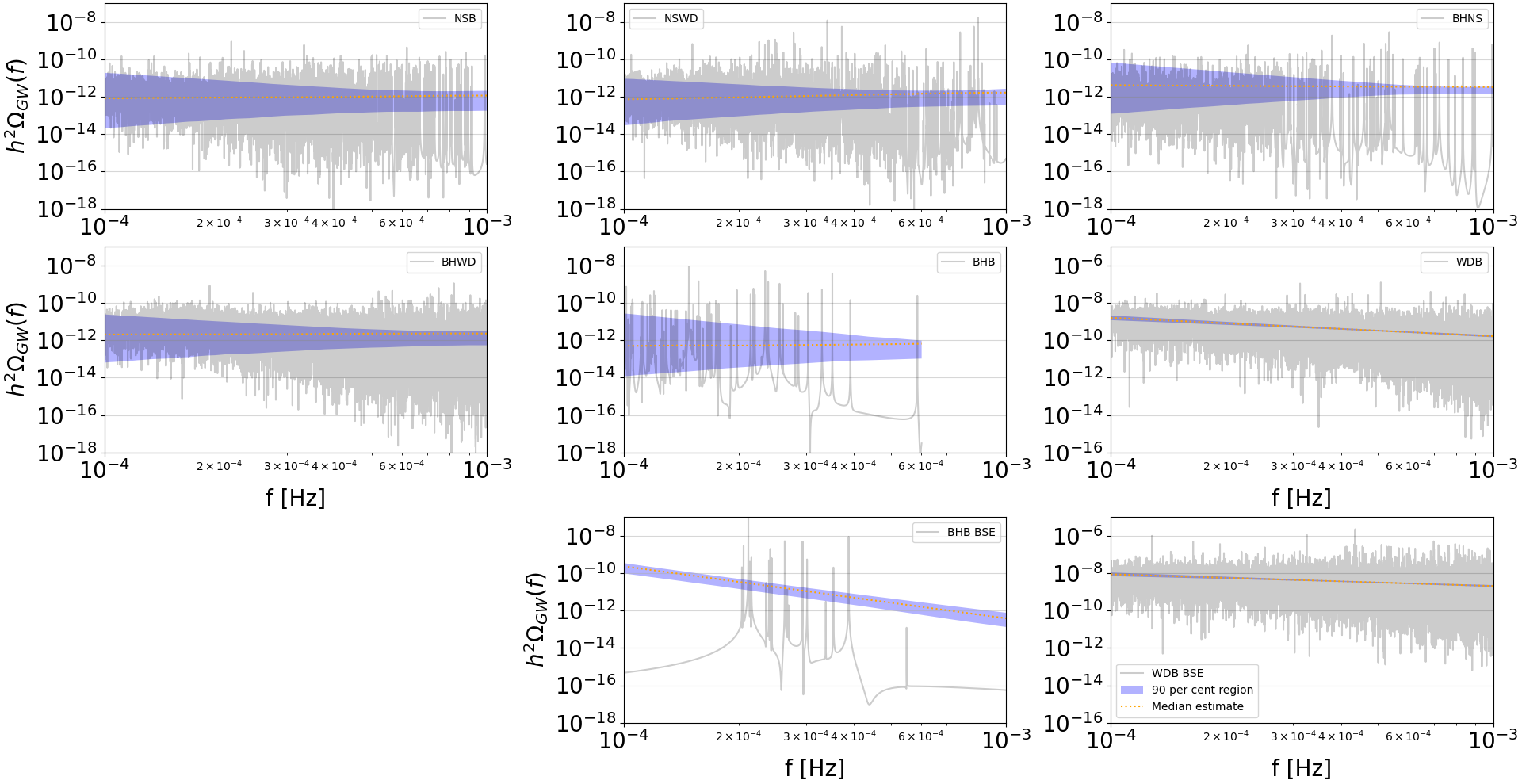} \\
    \vspace*{5mm}
    \includegraphics[width=1.8\columnwidth]{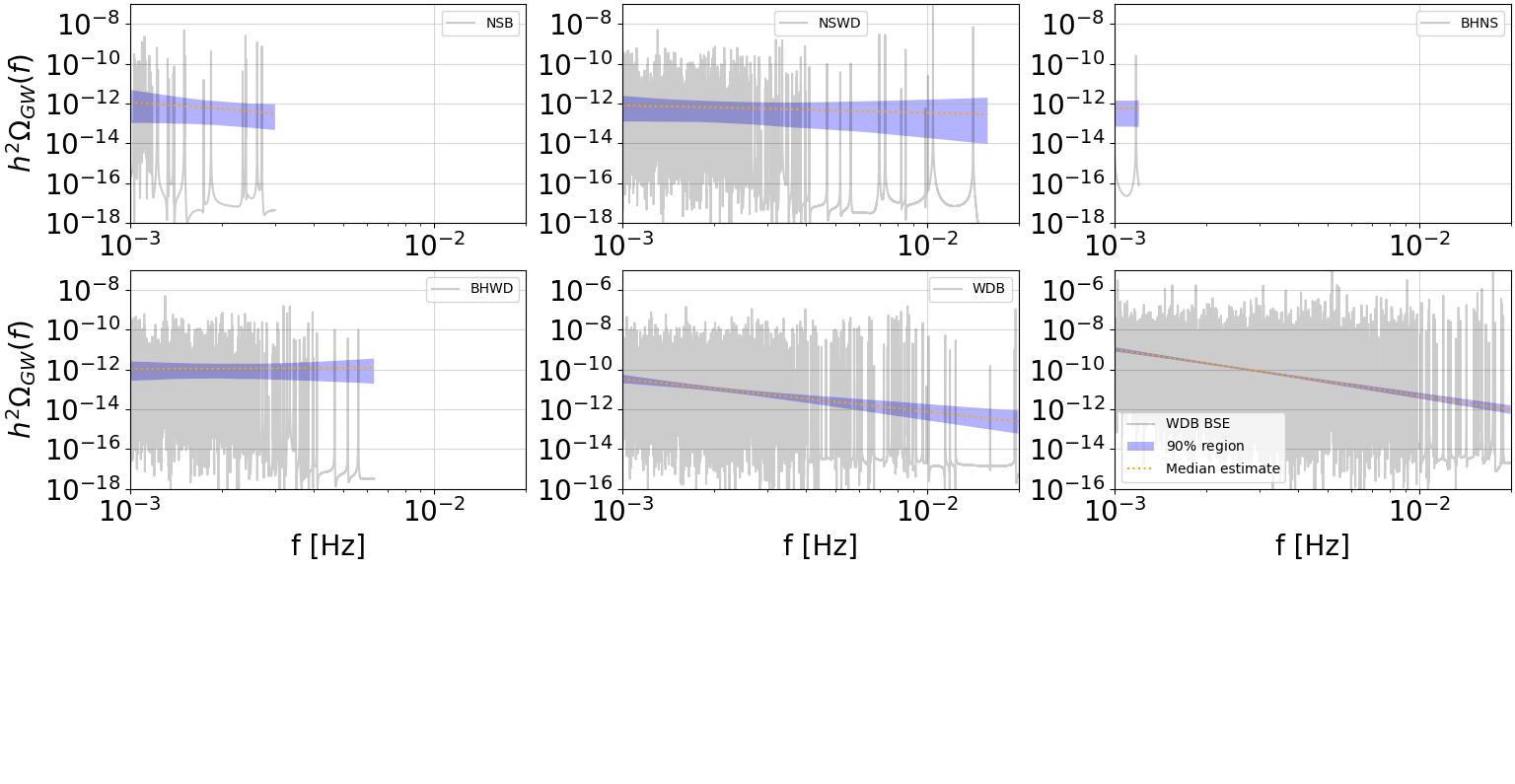}
    \vspace*{-20mm}
    \caption{Bayesian parametric single PL models for estimating ESD at frequency $\leq 1$~mHz (top) and $>1$~mHz (bottom). The grey lines represent the periodogram of the GW signal. The orange lines represent the mean ESD estimates, and the HDI is taken as 90 per cent shaded in blue.}
    \label{fig:estimate_highlow}
\end{figure*}

Next, we separately illustrate the outcomes of our Bayesian single PL models for estimating the ESDs at frequencies above and below 1~mHz in Figure~\ref{fig:estimate_highlow}. Compared with the estimates over the full LISA frequency range of 0.1~mHz to 30~mHz, the only difference here is that we assume the priors for $f_T$ and $f_b$ are $U(10^{-4}, 10^{-3})$ and $U(10^{-3}, 3\times 10^{-2})$ at frequencies below and above 1~mHz respectively. As shown in the bottom figure, at frequencies above 1~mHz, the 90 per cent regions are narrower compared with signals below 1~mHz; again, this is because the signals are stronger in the higher frequencies. A wider 90 per cent region also corresponds to less signal present.

We found that the mean and the median ESD estimates are similar in all Galactic binary populations. Therefore, we only plot the median estimates (in orange), seen in Figure~\ref{fig:estimate_highlow} (top). The median estimates are flat and more uncertain, closer to 0.1~mHz. This larger uncertainty is because the signals are weaker near 0.1~mHz. Finally, compared with the ESD estimates in Figure~\ref{fig:estimates}, Figure~\ref{fig:estimate_highlow} does not show a significant difference.

We observe that the single PL model proves to be less than ideal, as it does not capture spectral density features towards the higher frequency range within the LISA frequency band, i.e., fewer observations and more considerable uncertainties or larger HDI region. This limitation at higher frequency stems from several Galactic populations within BPASS, such as NSB, BHWD and WDB populations, which have fewer systems with high frequencies. Conversely, at lower frequencies, the uncertainties are smaller or narrower blue-coloured 90 per cent region with more observations.

\subsection{Total Galactic population results}
\label{5.2}

The signal that LISA will detect includes all Galactic binary populations. We combine the noise strain with all Galactic binary populations and estimate the ESD using our Bayesian parametric models. Our estimated single PL of total BPASS Galactic binary GW signal has $\alpha = 4.9^{+0.1}_{-0.1} \times 10^{-9}$ and $\beta = -2.33 ^{+0.04}_{-0.04}$, and we estimate the total Galactic binary population has the background energy density $h^2 \Omega_{\rm GW} = 1.1 ^{+0.2}_{-0.2} \times 10^{-9}$ at frequency 3~\text{mHz} in Table~\ref{tab:5}, which is the highest amongst the three model predictions. The corner plots for one MCMC estimation of all three GW ESD functional forms are presented in Figures~\ref{fig:pl_cornor}, \ref{fig:broken_cornor} and \ref{fig:peak_cornor}. These results show that the broken PL is less ideal as the uncertainties of the posteriors are large.

\renewcommand{\arraystretch}{2}
\begin{table}
  \centering
  \caption{Estimated mean ESD $h^2 \Omega_{\rm GW}$ at 3 $\text{mHz}$ of total BPASS Galactic binary population. The mean estimates are accompanied with the HDI is taken as 90 per cent.}
  \begin{tabular}{|c|c}
        Model             & $h^2 \Omega_{\rm GW}$ at 3 $\text{mHz}$\\
        \hline
        Single PL         & $1.1 ^{+0.1}_{-0.1} \times 10^{-9}$ \\
        Broken PL         & $3.5 ^{+0.3}_{-1.2} \times 10^{-10}$ \\
        Single-peak       & $1.5 ^{+0.4}_{-0.4} \times 10^{-9}$
    \label{tab:5}
  \end{tabular}
\end{table}

\begin{figure}
    \centering
    \includegraphics[width=\columnwidth]{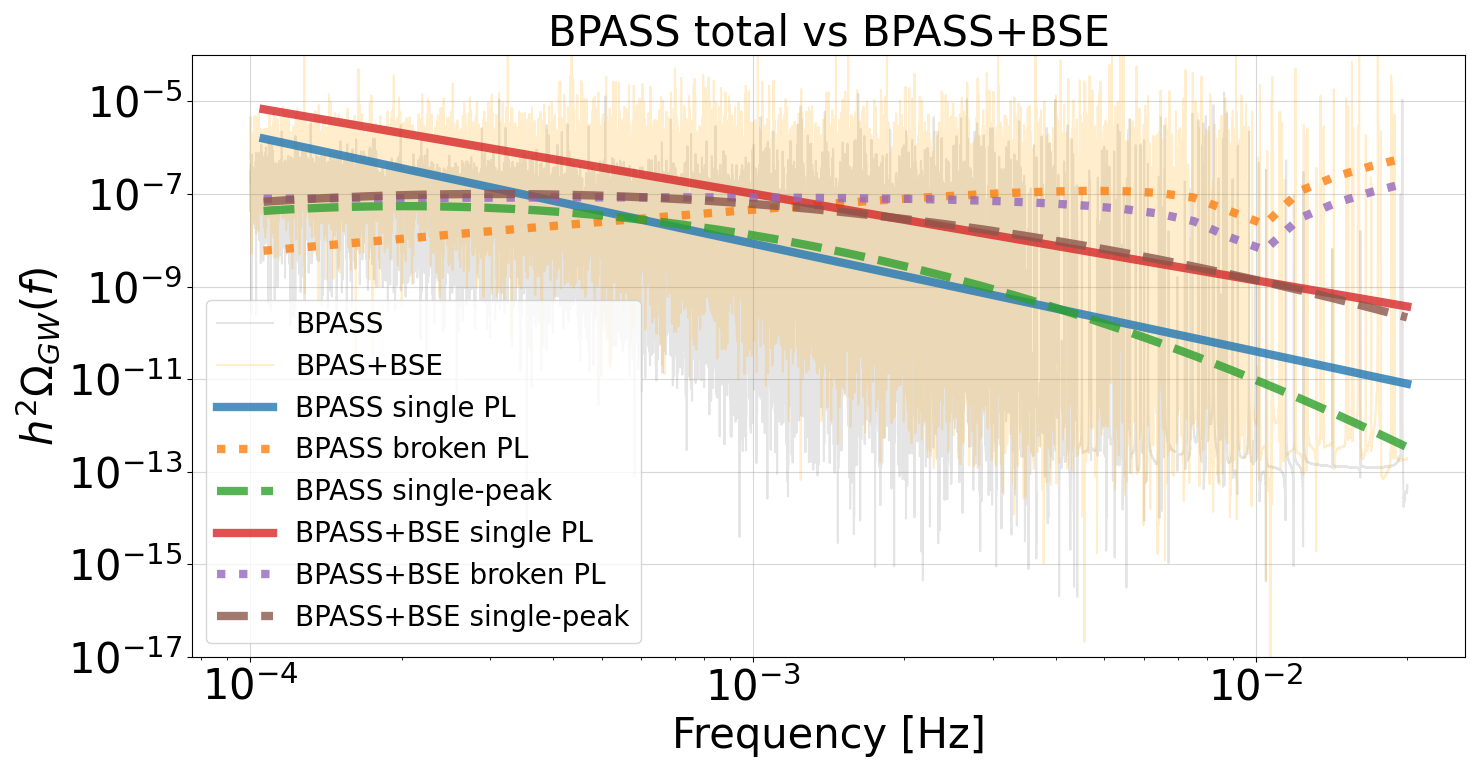} \\
    \caption{ Mean parametric model estimate of total BPASS Galactic binary population (in grey) vs BPASS + BSE population (in orange). Single PL models are represent in solid line, broken PL models are in dotted lines and the dashed lines represent the mean estimate of the single-peak model fits.}
    \label{fig:BPASSvsBSE}
\end{figure}

We compare the ESD estimates of the total BPASS Galactic population with a total BPASS+BSE Galactic population, where we replaced the BPASS BHB and WDB populations with those from BSE in \citet{2018MNRAS.480.2704L, 2019MNRAS.490.5888L}. We plot the posterior mean of the ESD for the three different models in Figure~\ref{fig:BPASSvsBSE}. The reason for replacing the BPASS WDB and BHB population with the BSE WDB and BHB population is to see how different WDB and BHB populations change overall estimations.

Since the BSE WDB population has a stronger ESD, we see that the total BPASS+BSE (in orange) is greater than BPASS (in grey), with more power(signal) present above 10~mHz. The difference is because the WDB population contributes more to the total signal, which is consistent with our recent work \citep[][]{2024arXiv240520484T}. Overall, the BPASS+BSE population has a higher mean single PL estimate (in red) than the BPASS population (in blue). The BPASS+BSE population has a higher mean broken PL estimate $< 1$~mHz (in purple) but a lower mean broken PL estimate $> 1$~mHz than the BPASS population (in orange). In addition, the BPASS has a lower mean single-peak estimate (in green) than BPASS+BSE (in brown) overall, but the differences increase in higher frequencies. 

\section{Discussion}
\label{sec:6}

We start with a review of previous studies that performed ESD estimations. The SGWB generated by NSBs and BHBs, as detected by the LVK, falls within the frequency range most sensitive to SGWBs, approximately 25~Hz. Currently, assuming a flat background, the expected value based on theoretical considerations for this background's energy density parameter is $\Omega_{\rm GW}=8.9^{+12.6}_{-5.6} \times 10^{-10}$ (unit: per log-frequency), with amplitude $\alpha = 5.4 \times 10^{-12}$ and slope $\beta = 2/3$ \citep{2019PhRvD.100f1101A}, and $\Omega_{\rm GW} \approx 10^{-11}$ by \citet{2019PhRvD.100f3004C}.

In addition, we list a few predictions for the Galactic BHB population ESD made by other studies. \citet{2019ApJ...871...97C} estimated that the background energy densities at 3~mHz are $\Omega_{\rm BHB}=2.7^{+2.8}_{-1.6} \times 10^{-12}$ (with 90 per cent Poisson error bounds) come from stellar origin BHBs. The same study also estimated the background energy density at 3~mHz of primordial origin BHBs is $\Omega_{\rm BHB}=4.3^{+5.9}_{-3.0} \times 10^{-12}$. The studies by \citet{2016PhRvL.116w1102S, 2016PhRvD..94j3011D, 2020MNRAS.493L...1C, 2021PhRvD.103d3002P, 2024PhRvD.109d2001M} estimated $h^2\Omega_{\rm BHB}$ to be in a range of [$10^{-13}, 10^{-11}$] at 3~\text{mHz}, where h = 0.67 is a constant. In addition, on average, stellar origin BHBs are also estimated to have $h^2 \Omega_{\rm GW} \in [6.65, 11.5] \times 10^{-13}$ by \citet{2023JCAP...08..034B}. 

As for the NSB population, one study by \citet{2019ApJ...871...97C} estimated that $\Omega_{\rm NSB}=1.7^{+3.5}_{-1.4} \times 10^{-12}$, with 90 per cent Poisson error bounds, for population I/II  stars. A more recent study by \citet{2021PhRvD.103d3002P} gave an estimate of $\Omega_{\rm NSB} < 10^{-13}$ at 4~\text{mHz}.

Another population that has been studied is the Galactic WDB population for LISA. The study by \citet{2021MNRAS.508..803B} gave an estimate of SGWB of astrophysical origin in the scale of $\Omega_{Astro} \approx 10^{-12}$, and Galactic WDBs in the scale of $\Omega_{\rm WDB} \approx 10^{-9}$. In addition, \citet{2023arXiv230812437R} suggested that $\Omega_{\rm WDB} \approx 10^{-11}$ at one~\text{mHz} in the Large Magellanic Cloud (LMC), and this value should be more significant in the Galactic WDB population.

Our analysis found that there are differences between different Galactic binary population ESDs. This is expected as these populations exhibit different physical characteristics, i.e.,eccentricities, masses, and orbital frequencies, contributing to the overall GW signal ESD. 

When considering various Galactic binary populations, such as NSB, NSWD, BHNS, and BHB systems, we observed minimal to no expected GW signals above 10~mHz. The GW signals above 10~mHz originate predominantly from WDB systems, as shown in this study. This finding complements the study by \citet{2019PhRvD.100f1101A}, which suggested that BHNS systems in the range of 2 to 10~mHz contribute to the guaranteed astrophysical components in the LISA frequency band. While their study focuses on extragalactic populations, our results indicate that in our simulated Galactic populations, no BHNS systems are present above 2mHz. This is due to our population synthesis code, which predicts that our more massive BH evolve quickly and move beyond the LISA frequency range.

Apart from the dominant WDB Galactic population, the other guaranteed GW astrophysical component is expected to result from BHNS Galactic binaries at frequencies between 2 and 10~mHz \citep[][]{1987ApJ...323..129E, 1997CQGra..14.1439B, 2022ApJ...937..118W}. Instead, we only observe contributions from NSWD, BHWD, and WDB Galactic binary populations within this frequency regime. As shown in Figure~\ref{fig:estimates}, the maximum frequency of an object decreases in the order of WDB, NSWD, BHWD, NSNS, BHNS and BHB; this is in order of increasing chirp mass for the binaries in question; this is because higher chirp masses mean the systems evolve quicker especially, in the orbital periods of tens of minutes to one minute, where the more massive systems evolve through these regions even more quickly than the lower mass systems. Therefore, it is no surprise that we do not see NSB, BHNS, or BHB with these higher frequencies since they have all merged.

One study by \citet{2017ApJ...851L...4C} suggests that the BHWD system's mass transfer is unstable, leading to fewer systems. In our simulations, we observe that the Galactic BHWD population consists of 2,427,133 BHWD systems (see Table~\ref{tab:numbersource}), which represents only 0.85 per cent of the total Galactic binary systems in the LISA frequency band. Consequently, we expect LISA to detect a low number of BHWD systems. As reported in \citet{2023MNRAS.524.2836V}, uncertainties persist in our predictions due to the notable signal strength emanating from the BHWD and BHNS populations. Our findings indicate that if we subtract the WDB population, the BHWD systems may significantly overshadow the ESD of the remaining Galactic population. Consequently, the influence of other Galactic binary populations weakens noticeably below the frequency threshold of 10~mHz.

Furthermore, we compared the estimated mean ESD of binary populations from different models and provided predictions for total Galactic binary GW signal parameters under various model assumptions. However, we encountered limitations in accurately estimating noise parameters using parametric models with the broken PL model. The variations in ESD estimates across frequency ranges underscore the stochastic nature of Galactic binary population dynamics and the complexities of modelling their GW signals.

When we simulated the total signal for LISA, we assumed that the noise could be modelled with a finite number of parameters; in our case, they are P and A. However, the real LISA noise is likely to be more complex. The noise ESD might have a different shape from what has been defined in this study, the probability distribution of the noise might not be Gaussian \citep[][]{2024arXiv241014354K}, and the noise might not be stationary, which will lead to correlations between frequencies \citep[][]{2024PhRvD.109d2001M}. However, we have yet to explore these aspects in this study.

Our attempt to improve the ESD estimate by splitting the frequency into lower and higher frequencies provided valuable insights into different Galactic populations. Our next step is to design more flexible models that can capture the stochastic nature of the signals \citep[such as the studies by][]{2019JCAP...11..017C, 2020PhRvD.102h4062E, 2021MNRAS.508..803B, 2024PhRvD.109h3029P}. In this study, we investigated Galactic binary populations where both stars are stellar remnants. Therefore, an extension of this research would involve studying Galactic binary populations where one star is in transition into a compact object such as a BH, NS, or WD, i.e., a main sequence star. Often, these stars can be detected as X-ray binaries if the compact object is a BH. These X-ray binaries are of great interest to the community as they could be detected with X-ray observations, contributing to the multi-messenger astronomy field.

Additionally, expanding this work could entail the prediction of extra-galactic LISA sources employing BPASS, consequently enabling the prediction of the comprehensive astrophysical SGWB, even though these SGWB would not likely to be observed by LISA \citep[][]{2003MNRAS.346.1197F, 2024arXiv240710642H, 2024A&A...683A.139S, 2025arXiv250618390B}.

Another extension we plan to explore is the development of non-parametric methods for the noise and SGWB, so-called agnostic methods which are not constrained by assuming fixed ESD functional forms \citep[such as][]{2024PhRvD.109h3029P, 2023JCAP...04..066B}. However, separating the foreground from other backgrounds or noises will be challenging when using non-parametric methods for the Galactic foreground. To address this challenge, the T channel can be utilised as under certain conditions it is assumed to contain only noise, which can help constrain and separate the Galactic foreground more effectively \citep[][]{2024PhRvD.109d2001M, 2025arXiv250706300S}, thereby extending this study.

\section{Conclusions}
\label{sec:7}

In this study, we generated noise and GW signals from individual Galactic binary populations and the total Galactic binary populations within the LISA frequency range. We also conducted Bayesian estimation to compare and characterise the ESD of the Galactic binary populations derived from these binary populations. While keeping the noise the same for all models, we compared estimates from different GW ESD functional forms, such as single PL, broken PL, and single-peak. We found discrepancies between different models as well as between different Galactic binary populations.

We summarise our findings as follow:
\begin{enumerate}

    \item We predict a lack of BHNS signals above 2~mHz, which complements studies such as \citet{2019PhRvD.100f1101A}, which suggested that BHNS systems will be visible to LISA above 2~mHz. A similar reason as in point (i) can explain the lack of a system at a higher frequency range.

    \item We further predict LISA will be able to see GW signal originating from Galactic NSWD and WDB populations above 1~mHz as they are above the LISA sensitivity curve. The WDB result has been shown in our recent work \citet{2024arXiv240520484T} and globular clusters \citep[][]{2023MNRAS.524.2836V}. We conclude that the Galactic NSWD and WDB populations dominate the LISA band's total Galactic binary population signal. Therefore, LISA should be able to detect them.

    \item Based on our simulated BPASS populations, the ESD of the Galactic binary populations is unlikely to take the form of a single PL law. The ESD varies more at higher frequencies for populations with fewer observations. This variability suggests that a single PL law may not adequately capture the complexity of the ESD shapes. Therefore, we seek a more flexible model to describe the ESD shapes, which could involve multi-component models or non-parametric approaches. Such models allow us to better account for the stochastic nature of the signals and improve the accuracy of our predictions. 

    \item BPASS Galactic BHB and WDB population ESDs are different from that of BSE Galactic BHB and WDB populations; this is also predicted in \citet{2024arXiv240520484T}, in particular, BPASS has fewer WDB systems at frequency $> 10$~mHz, and more BHB systems at frequency $> 0.2$~mHz. As a result, the total LISA foreground ESD shape mostly depends on the sum of WDB and BHB populations at lower frequencies and on the WDB population at higher frequencies. Since there are only a small number of BHBs, it is possible to identify them, as they appear as individual peaks in the frequency domain, and subtract them from the total signal. We have not subtracted the BHB population in this study as we focused on the total population signal.

    \item When assuming the ESD is in the form of a single PL, we predict that the total BPASS Galactic binary GW signal has amplitude $\alpha = 2.0^{+0.2}_{-0.2} \times 10^{-8}$ and slope $\beta = -2.64 ^{+0.03}_{-0.04}$, and we estimate the total Galactic binary population has the ESD $h^2 \Omega_{\rm GW} = 1.1 ^{+0.2}_{-0.2} \times 10^{-9}$ at frequency 3~\text{mHz}. We also show that the amplitude $\alpha$ and slope $\beta$ are negatively correlated.

    \item When assuming a broken PL model, the total BPASS Galactic binary population ESD has $h^2 \Omega_{\rm GW} = 3.6 ^{+0.4}_{-1.7} \times 10^{-10}$ at frequency 3~\text{mHz}.

    \item When assuming a single-peak model, the total BPASS Galactic binary population ESD has $h^2 \Omega_{\rm GW} = 1.4 ^{+0.6}_{-0.6} \times 10^{-9}$ at frequency 3~\text{mHz}.

    \item As shown in Table~\ref{tab:model_comparison}, we compared different GW ESD functional forms to fit the simulated observation, and the fits show that the Galactic binary population ESDs are not all the same for different binary populations. We notice the range of the ESD increases at higher frequencies, and there seems to be a maximum ESD in all the binary populations, which has also been observed in clusters \citep[][]{2023MNRAS.524.2836V}.

    \item Also shown in Table~\ref{tab:model_comparison}, our estimates only improved slightly at lower frequencies when we separately estimated the ESD for lower and higher frequencies. The ESD estimates at the higher frequencies showed more significant variation than those below one mHz. 

\end{enumerate}

Finally, our simulation-based method provides a powerful and convenient tool for predicting the Galactic binary populations for LISA. However, accurately describing GW ESDs proves challenging due to the stochastic nature of the realistic data. In some populations and frequency ranges, the Galactic background may exhibit a 'popcorn-like' structure, necessitating alternative modelling approaches \citep[][]{2023PhRvD.107j3026L}. Therefore, more flexible modelling methods and realistic data informed by a better understanding of the underlying physics are required.

\section*{Acknowledgements}
We thank Sohan Ghodla for valuable discussions on BPASS models and Guillaume Boileau for useful discussions on GW signal simulations. We gratefully acknowledge support by the Marsden Fund Council from New Zealand Government funding, managed by Royal Society Te Apārangi, the University of Auckland and the University of Warwick for their continuing support. BPASS is enabled by the resources of the NeSI Pan Cluster. New Zealand’s national facilities are provided by the NZ eScience Infrastructure and funded jointly by NeSI’s collaborator institutions and through the Ministry of Business, Innovation \& Employment’s Research Infrastructure programme. URL: \url{https://www.nesi.org.nz}.

\section*{Data Availability}
The BPASS Galactic binary populations used in this article will be shared on reasonable request to the corresponding author. For more a general BPASS output regarding binary evolution please visit \url{https://bpass.auckland.ac.nz} or \url{http://warwick.ac.uk/bpass}.

\appendix

\section{Theoretical signal and Bayesian model simulation studies}
\label{sec:simulation}
We present the results of a simulation study in order to validate our Bayesian models. This section also includes supplementary plots from the main paper. We hope these extra plots provide further insights and a comprehensive view of our findings.

\subsection{Signals with sources SNR $>$ 7 removed}
\label{sec:7SNR}
\begin{figure*}
    \centering
    \includegraphics[width=2\columnwidth]{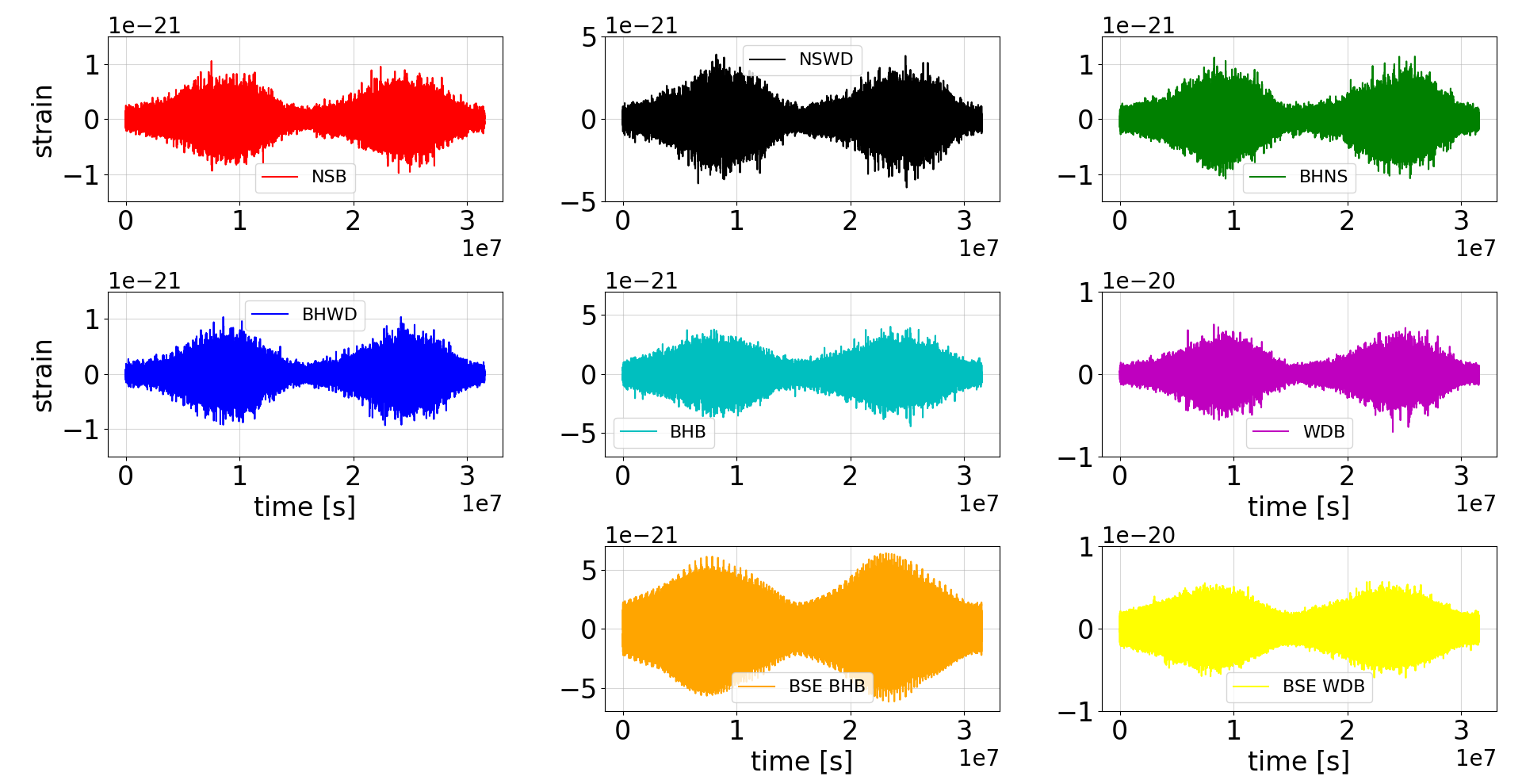}
    \caption{ Modulated signals over one-year LISA mission for a single channal A, signal frequencies ranging from 0.1~mHz to 0.1~Hz.}
    \label{fig:modulated_7snr}
\end{figure*}

\begin{figure*}
    \centering
    \includegraphics[width=2\columnwidth]{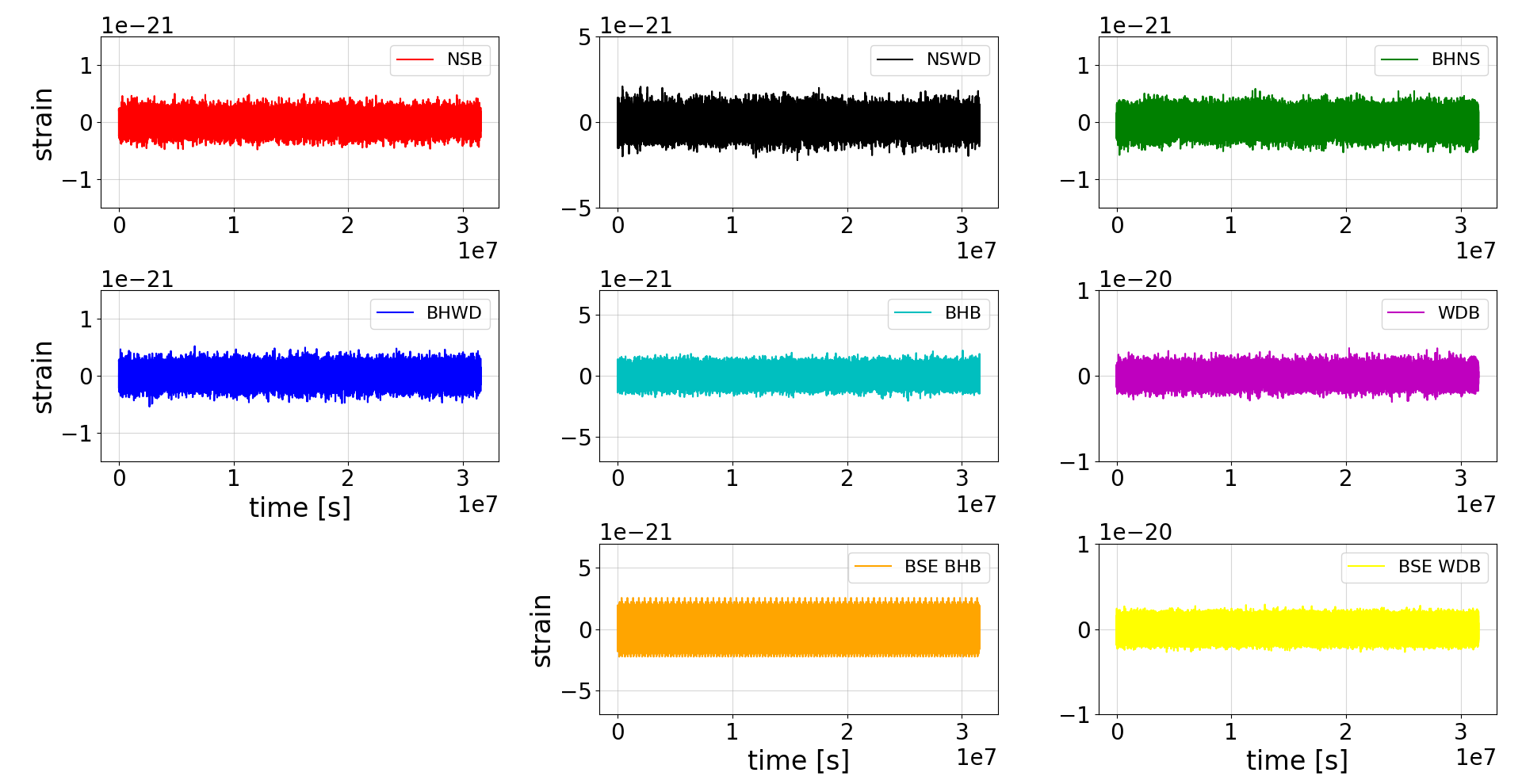}
    \caption{ Response function $\overline{F}_{A}(t)$ averaged signals over one-year LISA mission, signal frequencies ranging from 0.1~mHz to 0.1~Hz.}
    \label{fig:averaged_7snr}
\end{figure*}

The modulated and averaged signals weaken after sources with SNR $>$ 7 are removed, this is shown in Figures~\ref{fig:modulated_7snr} and \ref{fig:averaged_7snr}. We notice the modulation of the signals remains.

\subsection{Theoretical signals, Priors and Likelihood}
\label{sec:theoritical signal}

Table~\ref{tab:1} displays priors for single PL, broken PL and single-peak models and the theoretical `true' parameter values for generating the LISA data to test our Bayesian models. All prior distributions are relatively informative. The noise priors are set to be Gaussian centred at 15 and 3 for $P$ and $A$, respectively; the assumption is that both values are known within about 20 per cent \citep{2018CQGra..35p3001C}.

\begin{table}
    \centering
    \caption{Priors for single PL (top), broken PL (middle) and single-peak (bottom) test models.}
    \begin{tabular}{ccc}
         & \multicolumn{1}{c}{\textbf{Single PL}} \\
     \hline
        Parameter & Prior & Theoretical true value\\
        \hline
        P & Norm(15, 3) & 15 \\
        A & Norm(3, 0.6) & 3 \\
        $\alpha$ & HalfNorm($\sigma=10^{-10}$ ) & 5.4 $\times 10^{-12}$ \\
        $\beta$ & U(0, 1) & 2/3 \\
     \hline        
     & \multicolumn{1}{c}{\textbf{Broken PL}} \\
     \hline
        Parameter & Prior & True value \\
        \hline
        P & Norm(15, 3) & 15 \\
        A & Norm(3, 0.6) & 3 \\
        $\alpha$ & Norm( 6.4 $\times 10^{-12}$, $10^{-11}$ ) & 6.4 $\times 10^{-12}$ \\
        $\delta$ & Norm(-1/3, 1/2) & -1/3 \\
        $\gamma$ & Norm(2/3, 1/3) & 2/3 \\
    \hline        
     & \multicolumn{1}{c}{\textbf{Single-peak}} \\
     \hline
        Parameter & Prior & True value \\
        \hline
        P & Norm(15, 3) & 15 \\
        A & Norm(3, 0.6) & 3 \\
        $\alpha$ & HalfNorm( $\sigma=10^{-10}$ ) & 1 $\times 10^{-11}$ \\
        $\Delta$ & U(0.01, 2) & 0.2 \\
        $f_b$ & U($10^{-4}$, $10^{-2}$) & 0.003 \\    
    \end{tabular}
    \label{tab:1}
\end{table}

We use the Whittle likelihood approximation defined in Eq.~\ref{likelihood}. For posterior sampling, we use PyMC3 to draw 3,000 samples from the posterior and verify that this provides accurate estimates of the true parameters of the known ESD.

\subsection{Single power law}
\label{section:PL_post}

The simplest GW signal to reconstruct is a single PL, given in Eq.~\ref{PL}. We choose an input signal with `true' amplitude $\alpha = 5.4 \times 10^{−12}$ and slope $\beta$ = 2/3, which are motivated by the astrophysical SGWB from BHB and NSB populations \citep{2019PhRvD.100f1101A}.

\begin{figure}
    \centering
    \includegraphics[width=10cm]{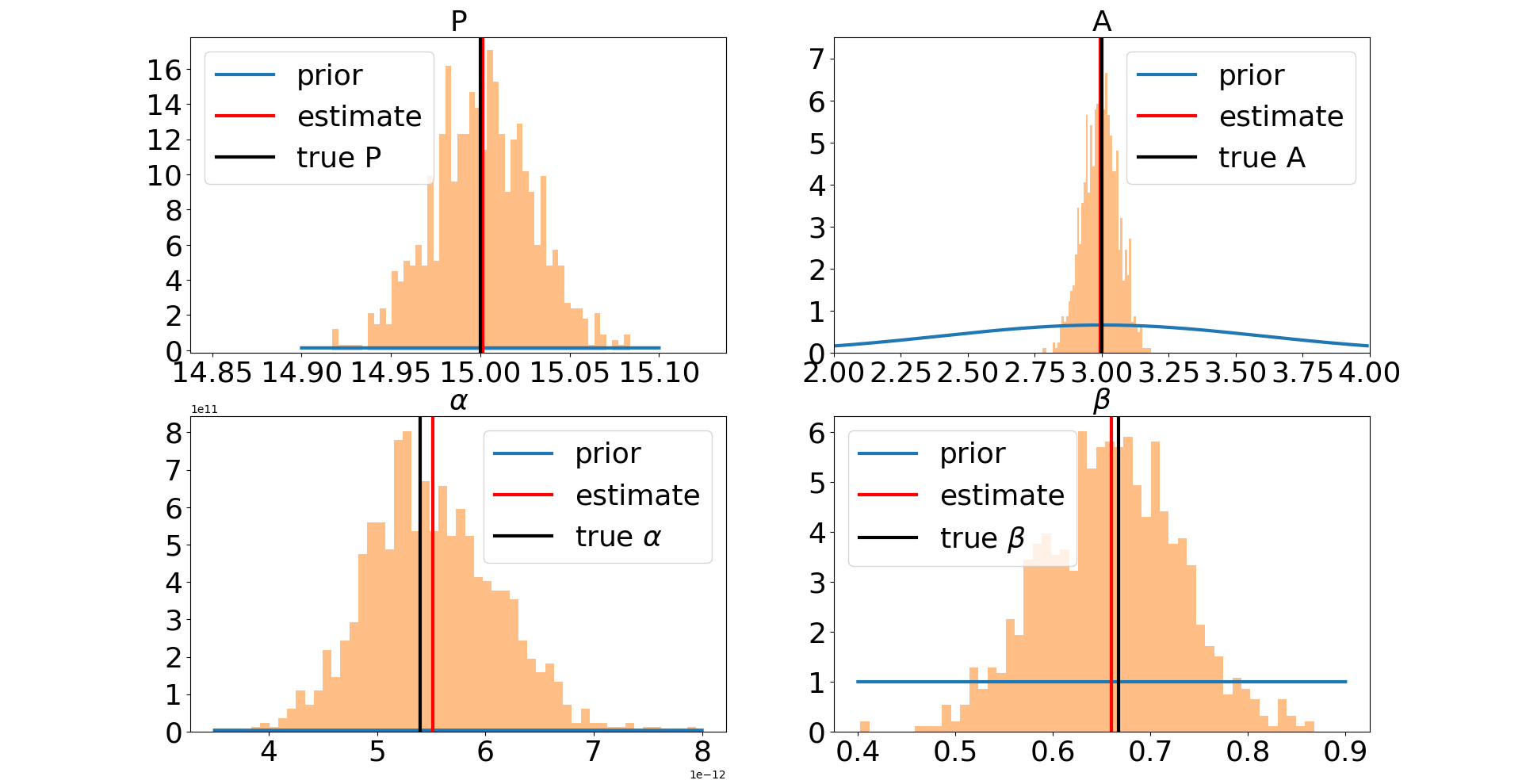}
    \caption{Distributions of simulation study results for 1,000 single PL parameters mean estimations.Black lines represent `true' parameter values, red lines represent mean estimate values and blue lines represent prior distributions.}
    \label{fig:simulation_pl}
\end{figure}

To set up and perform the simulation study, we generated the real and imaginary parts of the signal at each 99,900 frequencies. To account for randomness in our models and obtain a distribution of possible mean parameter estimates, we repeat the parameter estimation 1,000 times with different random seeds. The distribution of mean parameter estimates is shown in Figure ~\ref{fig:simulation_pl}. 

\begin{figure}
    \centering
    \includegraphics[width=8cm]{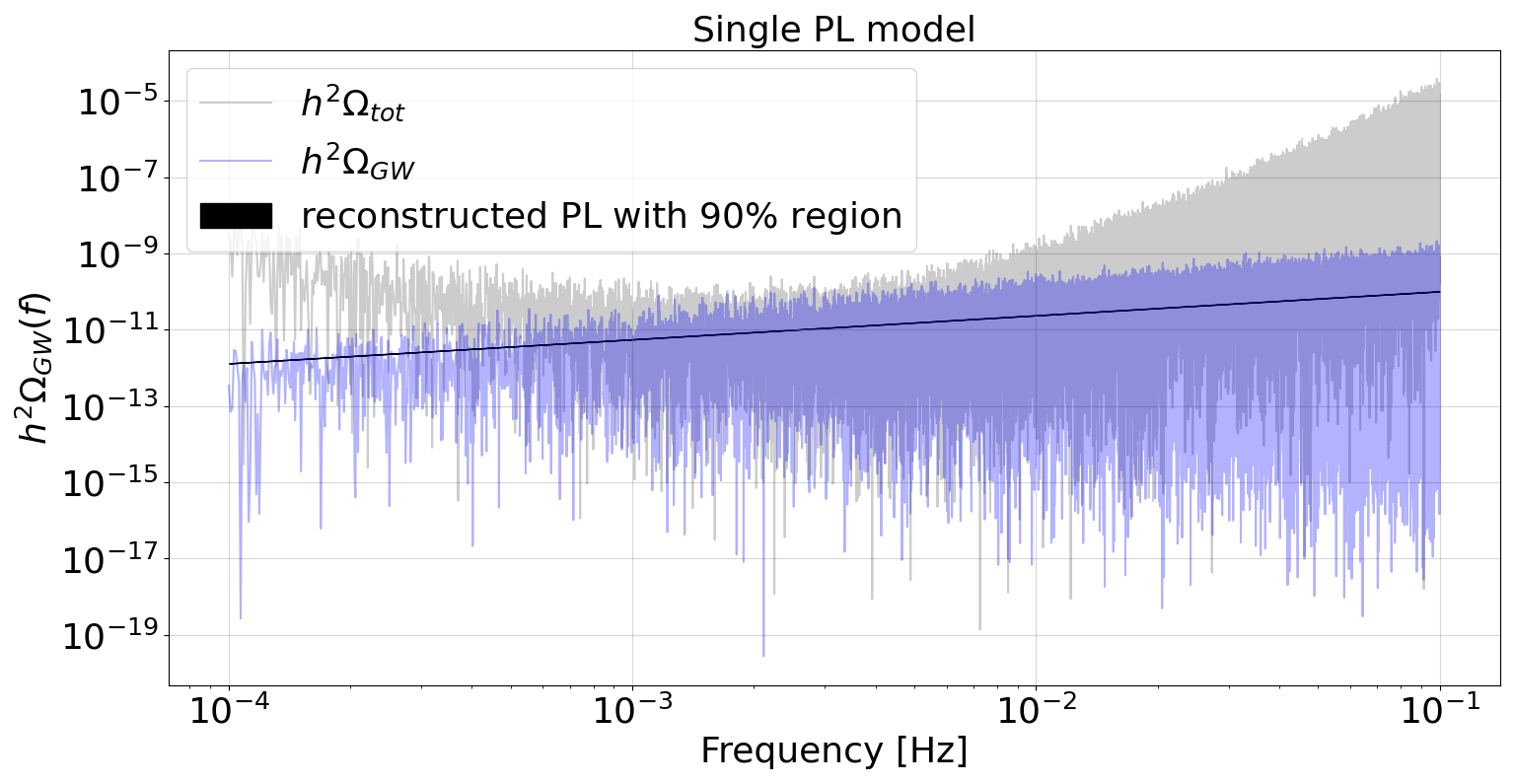}
    \caption{Two-parameter $\alpha$ and $\beta$ reconstruction single PL. The grey line represents periodograms of the Galactic SGWB of the total GW signal + noise (in grey) and GW signal only (in blue) overlaid by the estimated ESD (in black) for one simulation of the single PL model. ESD estimate with 90 per cent region.}
    \label{fig:pl_test2}
\end{figure}

\begin{figure*}
    \centering
    \includegraphics[width=\textwidth]{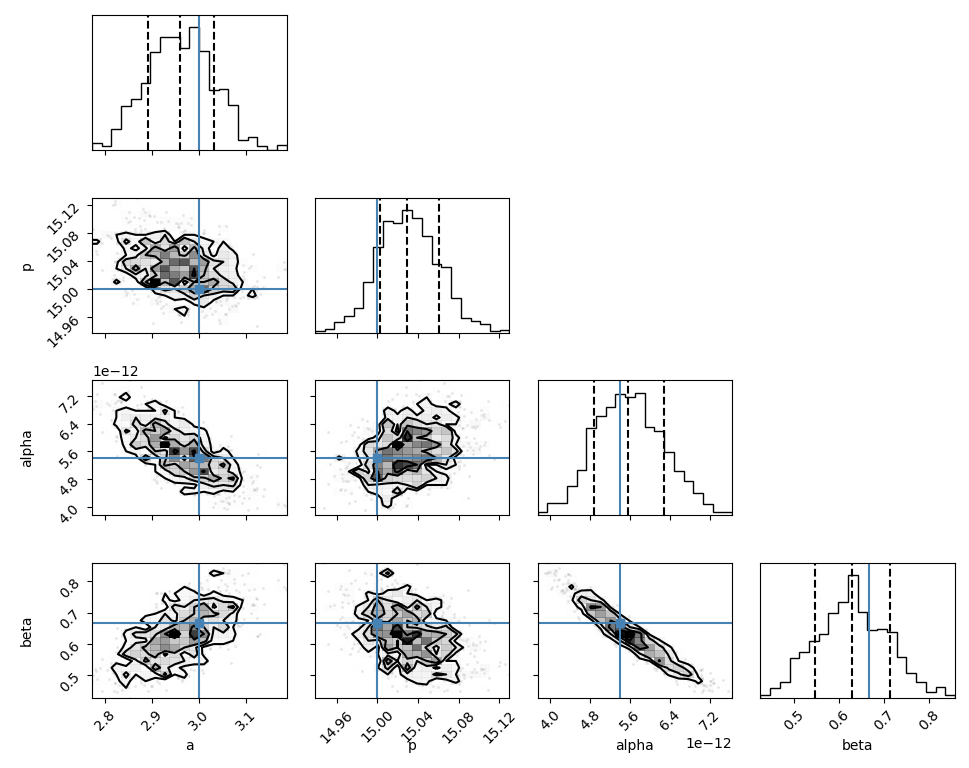}
    \caption{Corner plot of posterior distributions of four parameters based on single simulation. The results are for the LISA noise (parameters P and A) and single PL (parameters $\alpha$ and $\beta$). The blue lines are the ``true'' parameter values. The vertical dashed lines on the posterior distribution represent from left to right the quantiles [16 er cent, 50 per cent, 84 per cent].}
    \label{fig:pl_test1}
\end{figure*}

From one simulation, in Figure~\ref{fig:pl_test2}, we show the posterior mean of a single PL GW signal with a 90 per cent region in black, and the combined GW and noise ESD are in grey. The credible regions are tiny because we generated a very long time series. This trend is also seen in study \citep[such as][]{2018CQGra..35p3001C}. Corner plots of all model parameters $\alpha$, $\beta$, P and A are shown in Figure \ref{fig:pl_test1}, where the blue lines are the `true' parameter values. The vertical dashed lines on the posterior distribution represent, from left to right, the quantiles [16 per cent, 50 per cent, 84 per cent]. We see a negative correlation between the amplitude $\alpha$ and the slope $\beta$. Compared with the `true' parameter values in Table~\ref{tab:1}, the estimated values of all the model parameters are pretty close to the true values. The estimated ESDs and the corner plots demonstrate that we can estimate the ESD using our parametric models well. We note that the actual LISA signals will be more stochastic, and considerably more uncertainty will be expected.

We note that all parameter labels are interchangeable, such as the Greek letter $\alpha$ and the English spelling Alpha.
\subsection{Broken power law}
\label{section:broken_post}

Several studies \citep[such as][]{2019JCAP...11..017C} assumed their broken PL models have fixed central frequency, where they kept it fixed at $f_t = 0.0001$~Hz. We define a broken PL from Eq.~\ref{Broken} as a piece-wise linear function where the slope changes from $\gamma$ to $\delta$ at some switch frequency. This section lets the switch frequency $\rm f_T$ be 0.002~Hz. However, $\rm f_T$ is a parameter we estimate in Section~\ref{sec:4}.

\begin{figure}
    \centering
    \includegraphics[width=8.5cm]{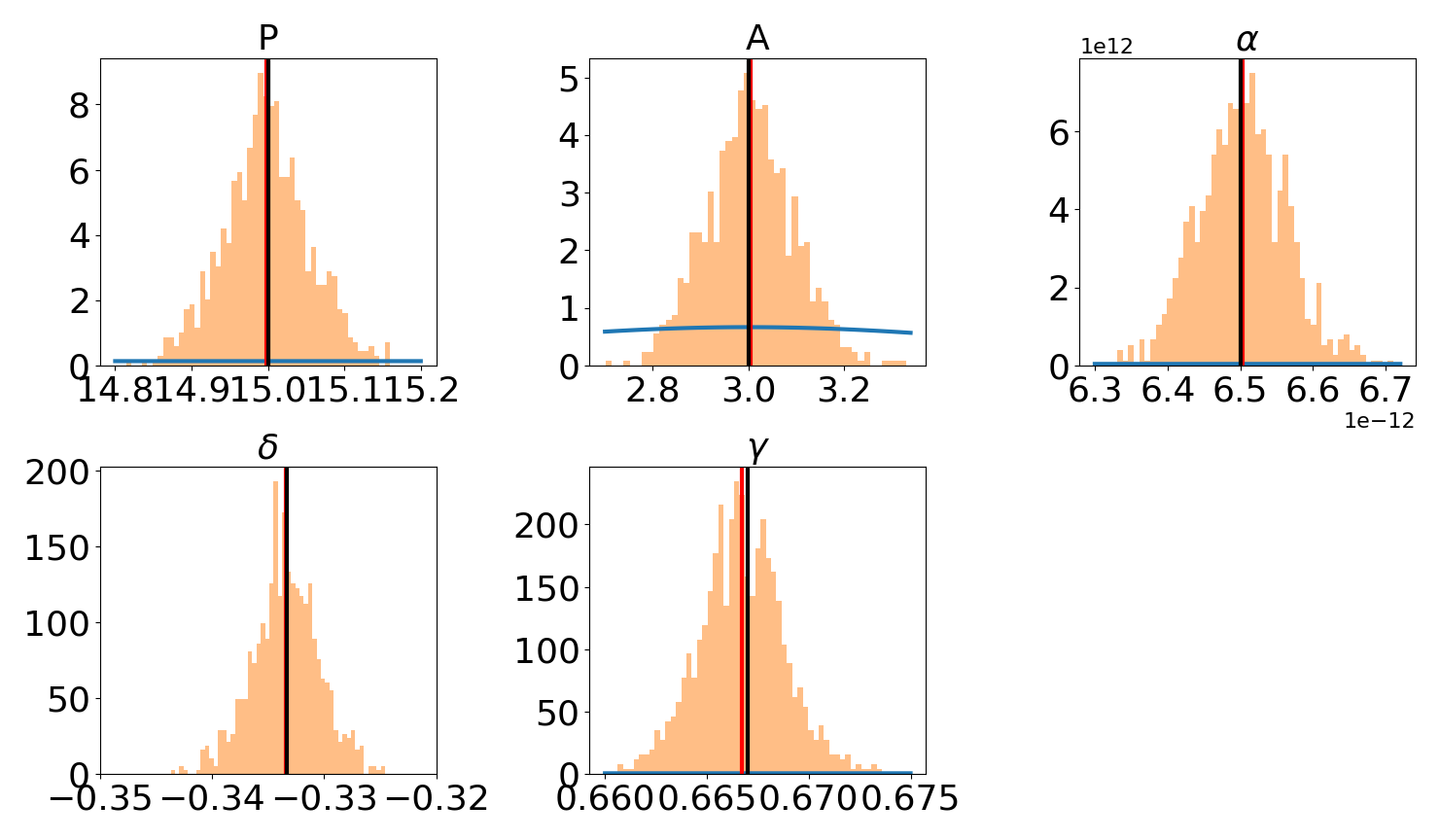}
    \caption{Distributions of simulation study results for 1,000 broken PL parameters mean estimations. Black lines represent `true' parameter values, red lines represent mean estimate values and blue lines represent prior distributions.}
    \label{fig:simulation_broken}
\end{figure}

Similar to the single PL simulation study, we repeat the parameter estimation 1,000 times with random seeds. The distribution of mean parameter estimates is shown in Figure~\ref{fig:simulation_broken}.

\begin{figure}
    \centering
    \includegraphics[width=8cm]{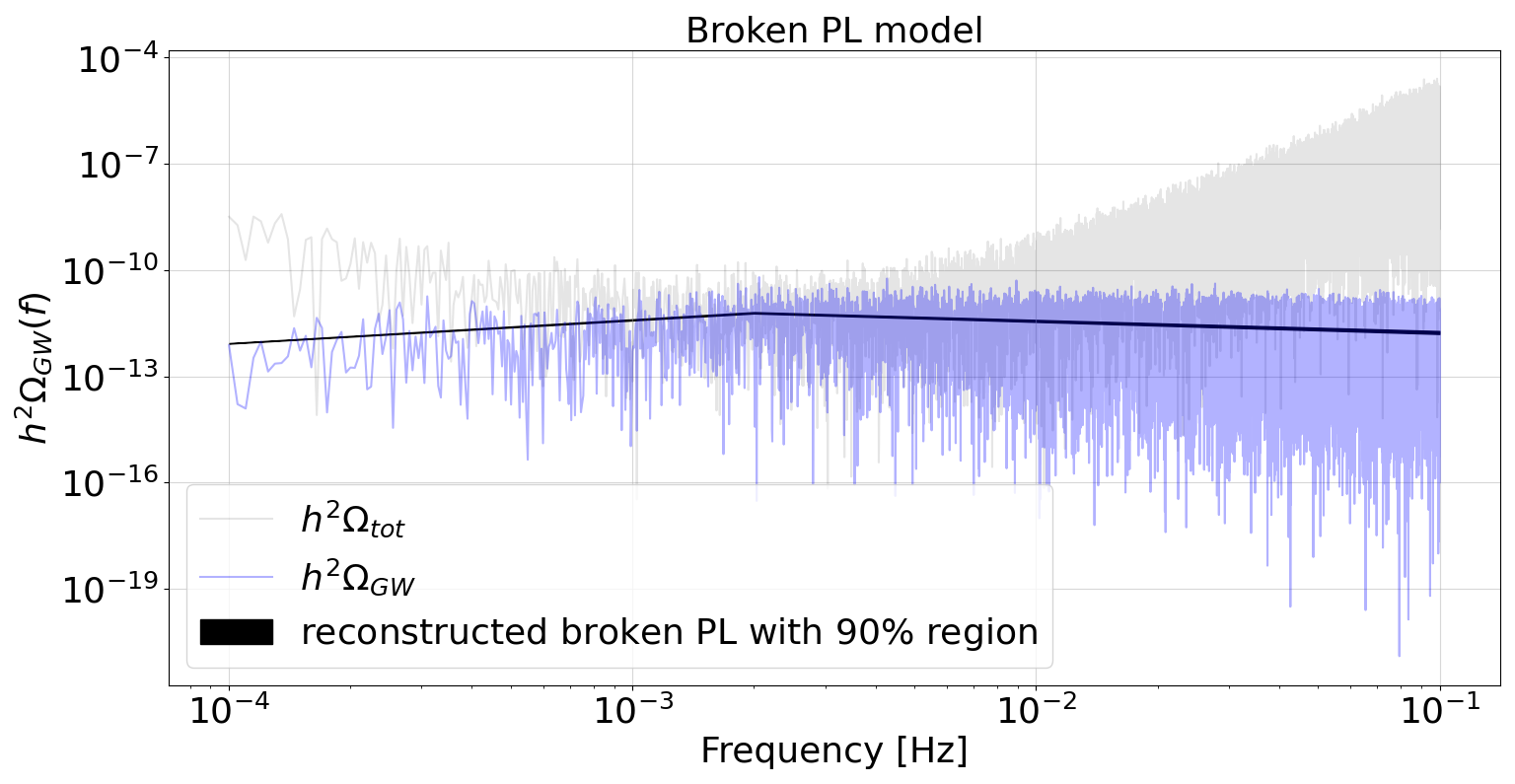}
    \caption{Three-parameter $\alpha$, $\delta$ and $\gamma$ reconstruction broken PL of a test run. The grey line represents the total signal of noise + GW signal, the purple line represents the GW signal only.}
    \label{fig:broken_test2}
\end{figure}

\begin{figure*}
    \centering
    \includegraphics[width=\textwidth]{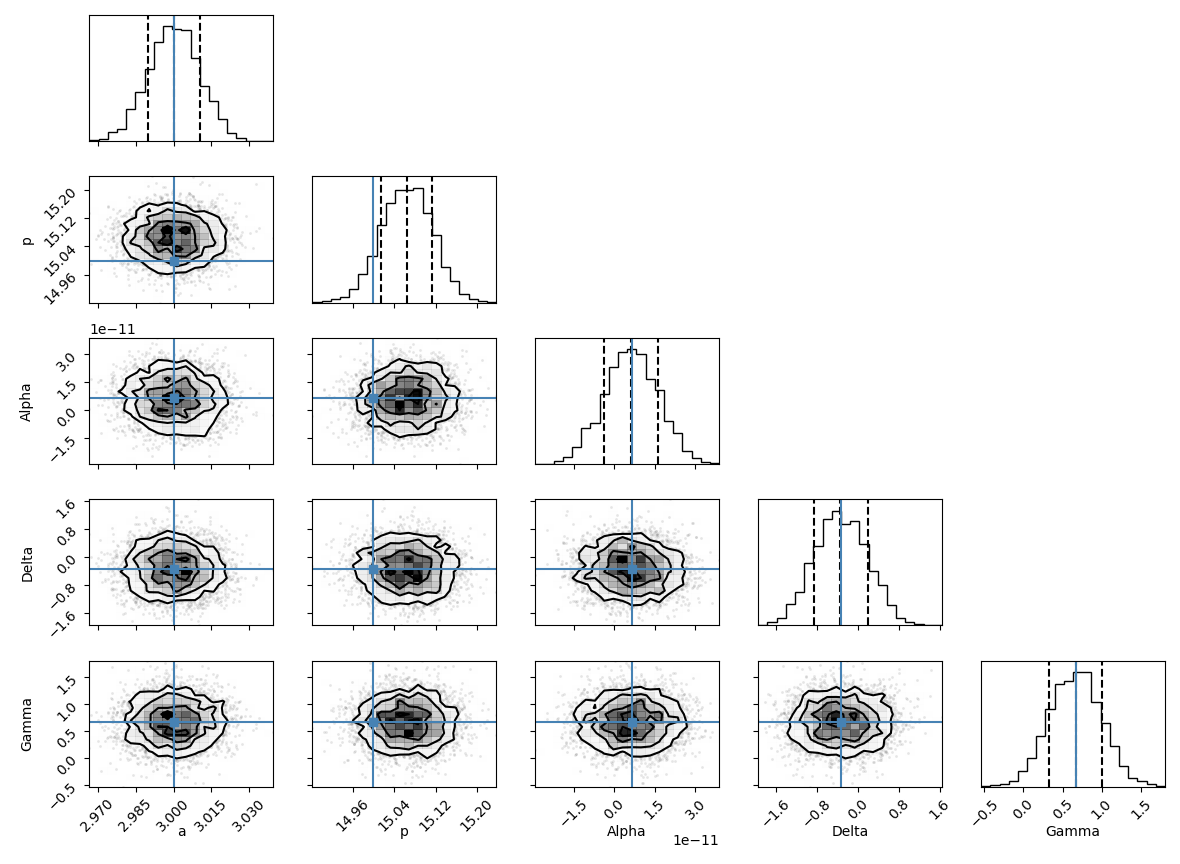}
    \caption{Corner plot for one MCMC estimation of five parameters. The results are for the LISA noise (parameters P and A) and broken PL (parameters $\alpha$ $\delta$ and $\gamma$). The blue lines are the `true' parameter values. The vertical dashed lines on the posterior distribution represent from left to right the quantiles [16 er cent, 50 per cent, 84 per cent].}
    \label{fig:broken_test_cornor}
\end{figure*}

Shown in Figure~\ref{fig:broken_test2}, the result from one simulation, we estimated the GW signal with a 90 per cent region in black, and the combined GW and noise ESD are in grey. A corner plot of all parameters used in broken PL is shown in Figure~\ref{fig:broken_test_cornor}, where the blue lines are the `true' parameter values. Like the single PL estimates above, the broken PL ESDs and the corner plot demonstrate that we can also estimate the ESD using our parametric models.

We do not see any correlation between parameters based on the broken PL corner plot in Figure~\ref{fig:broken_test_cornor}. Comparing with the `true' parameter values in Table~\ref{tab:1}, the estimated parameter values of the broken PL agree with the `true' values.

\subsection{Single-peak}
\label{section:peak_post}

\begin{figure}
    \centering
    \includegraphics[width=8.5cm]{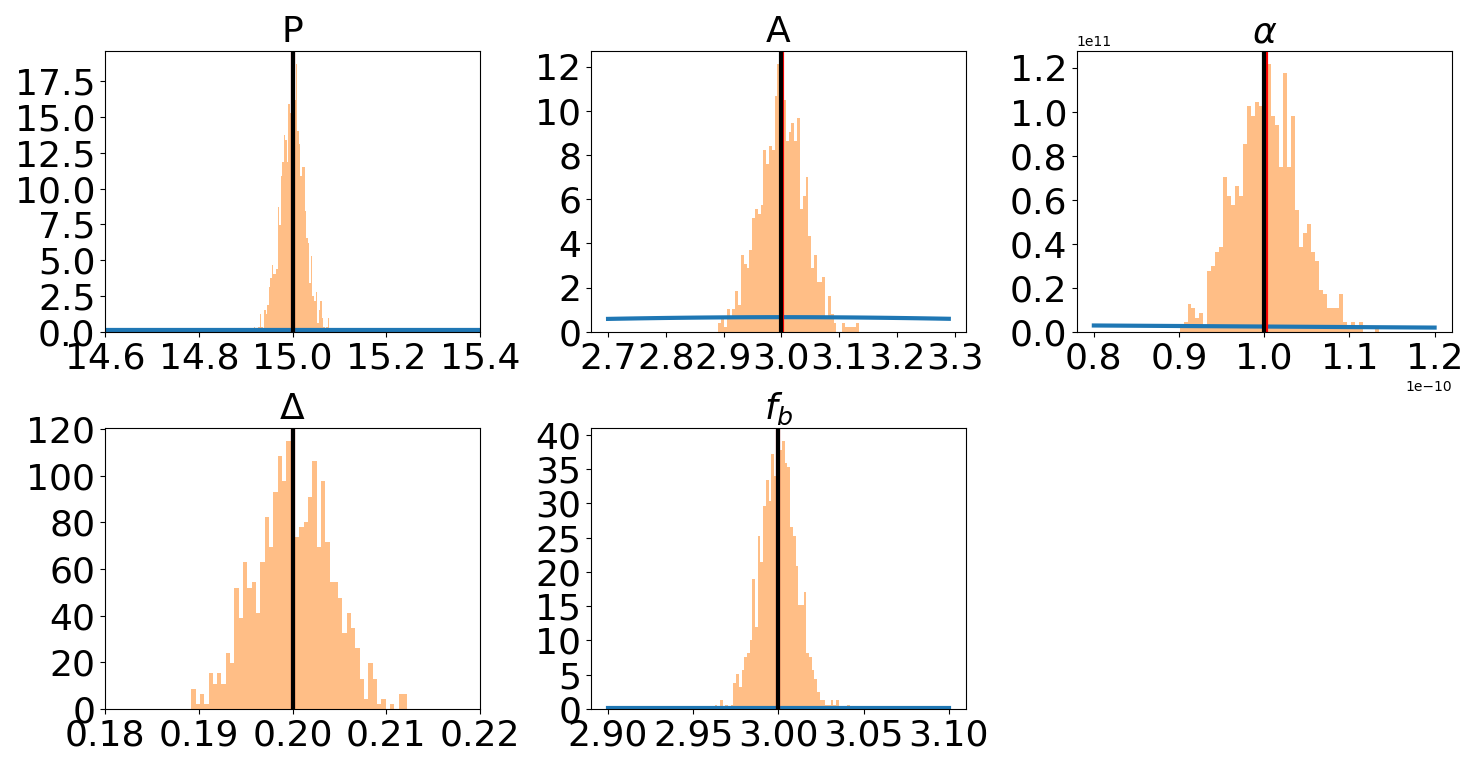}
    \caption{Distributions of simulation study results for 1,000 single-peak parameters mean estimations. Black lines represent `true' parameter values, red lines represent mean estimate values and blue lines represent prior distributions.}
    \label{fig:simulation_peak}
\end{figure}

Assuming the GW signal is a single-peak ESD shape, we generated the theoretical signal using the `true' parameter values based on the bottom panel in Table~\ref{tab:1}. In Figure~\ref{fig:simulation_peak}, we show simulation results of the distributions of mean parameter estimates. 

\begin{figure}
    \centering
    \includegraphics[width=8cm]{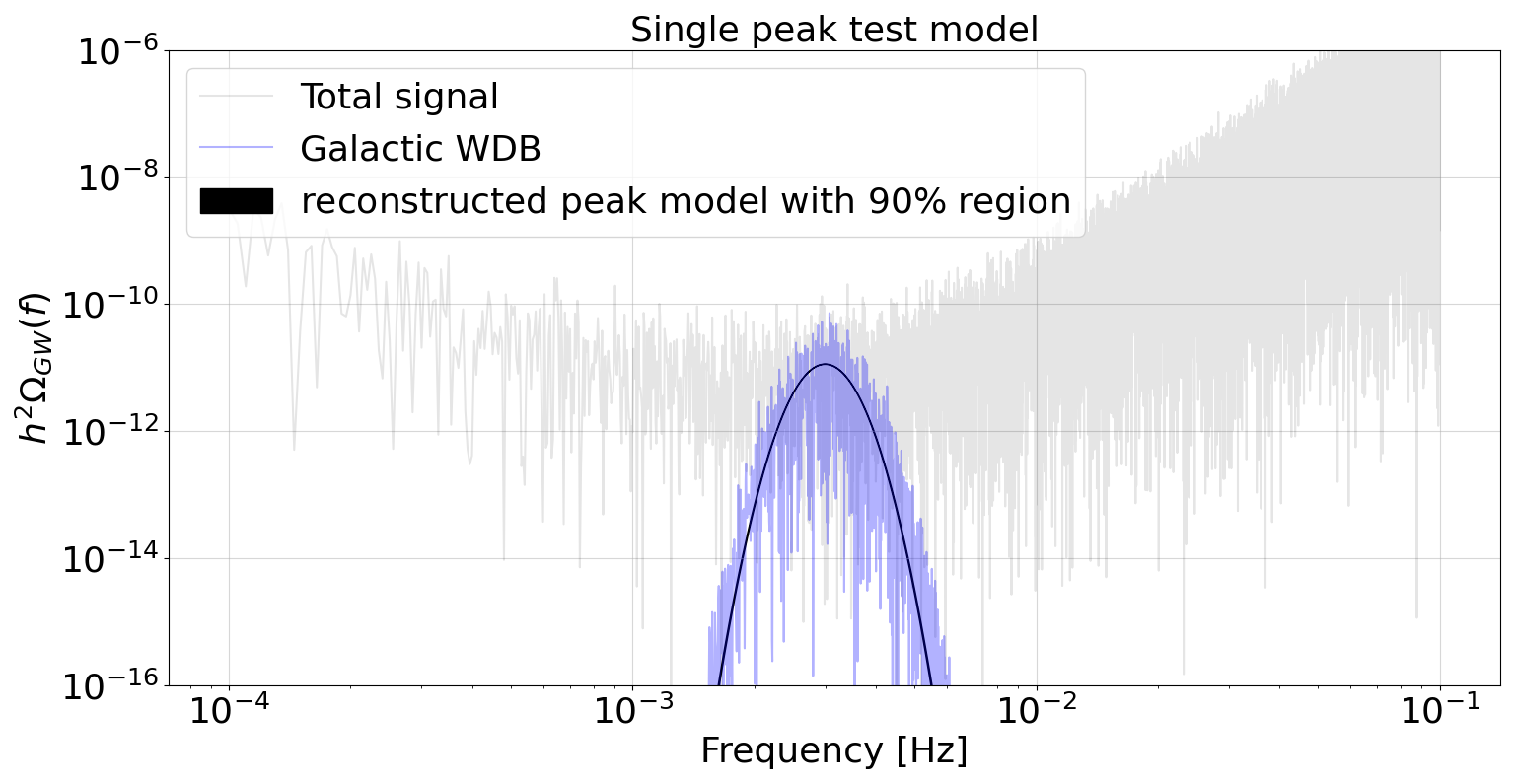}
    \caption{Three-parameter $\alpha$, $\Delta$ and $f_b$ reconstruction single-peak of a test run. The grey line represents the total signal of noise + GW signal, the purple line represents the GW signal only.}
    \label{fig:peak_test2}
\end{figure}

\begin{figure*}
    \centering
    \includegraphics[width=\textwidth]{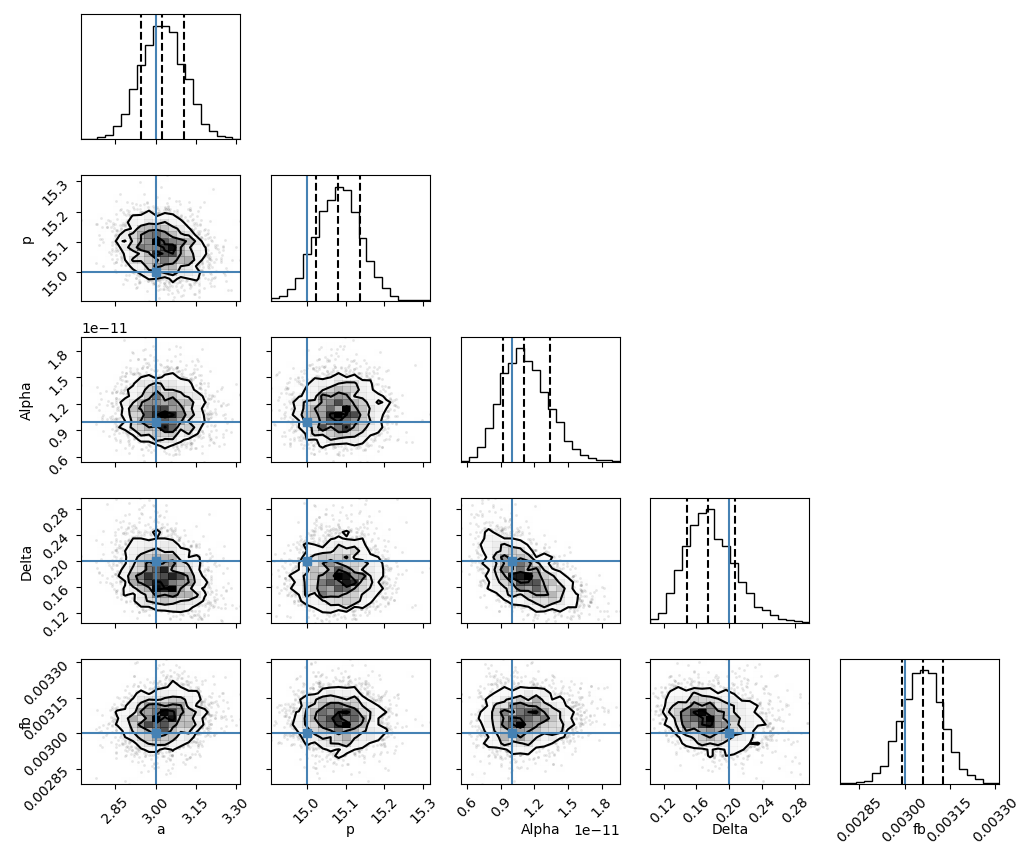}
    \caption{Corner plot for one MCMC estimation five parameters. The results are for the LISA noise (parameters P and A) and single-peak model (parameters $\alpha$ $\Delta$ and $f_b$). The blue lines are the `true' parameter values. The vertical dashed lines on the posterior distribution represent from left to right the quantiles [16 er cent, 50 per cent, 84 per cent].}
    \label{fig:peak_test_cornor}
\end{figure*}

For a single simulation, we estimated the GW signal with a 90 per cent black region; the combined GW and noise ESD are in grey for all GW functional forms, shown in Figure~\ref{fig:peak_test2}. Corner plots of model parameters $P$, $A$, $\alpha$, $\Delta$ and $f_b$ used in the single-peak models are shown in Figure~\ref{fig:peak_test_cornor}. The estimated ESDs and the corner plots demonstrate that we can estimate the ESD using our parametric models, as the estimated parameter values are close to the `true' parameter values.

Similar to the broken PL parameter estimates, we do not see any correlation between single-peak model parameters based on the corner plots in Figure~\ref{fig:peak_test_cornor}. Compared with the `true' parameter values in Table~\ref{tab:1}, the estimated values of the single-peak model parameters agree with the `true' values.

\section{Galactic binary population frequency distributions and ESD}

\begin{figure*}
    \centering
    \includegraphics[width=\textwidth]{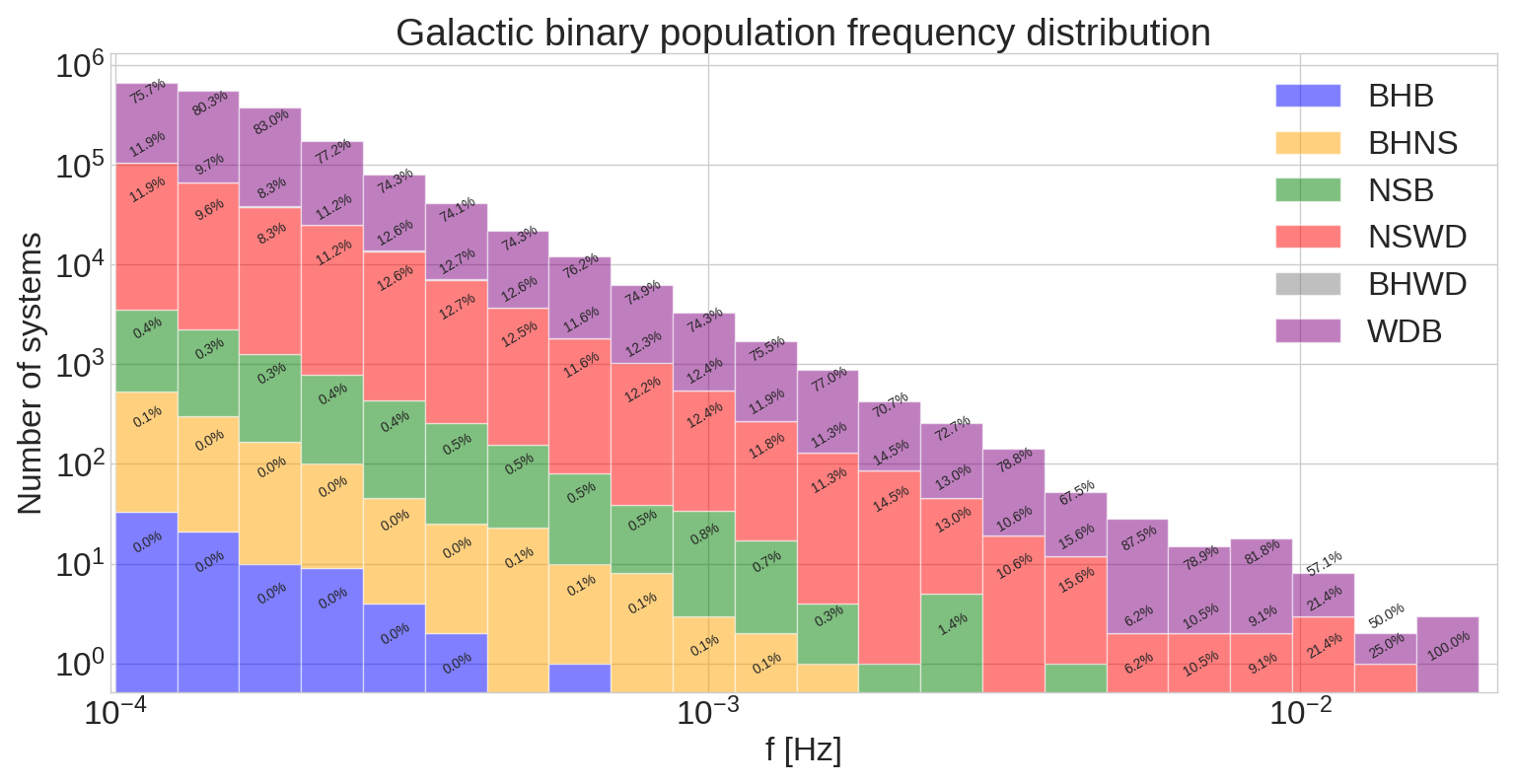} \\
    \caption{Frequency distributions of Galactic binary populations. }
    \label{fig:frequency}
\end{figure*}

The histogram of Figure~\ref{fig:frequency} shows the percentage of Galactic binary systems of different origins in different frequency bins. Where labels nsns and NSB are interchangeable, each frequency bin sums up to 100 per cent, and it is clear that binary with at least one WD dominates all frequency bins. This trend is reasonable as most of the stars in the Galaxy are WDs. Interestingly, BPASS predicts the second most numerous Galactic binary population is the BHWD Galactic binary population (in grey), and BHBs are only found in frequencies less than 1~mHz.


\section{No-U-Turn sampler}
\label{sec:4.2}

PyMC3 is a Python library for probabilistic programming, providing a flexible framework for building and fitting Bayesian statistical models. PyMC3 has been employed to aid many studies, such as the analysis of galaxy cluster data \citep[i.e.][]{2020MNRAS.498.4192F}, to study GWs from LIGO \citep[i.e.][]{2020PhRvD.102h4025V, 2021MNRAS.502.5576K}, and model planetary system dynamics \citep[i.e.][]{2021zndo...7191939F}. PyMC3 offers a range of MCMC algorithms for sampling from posterior distributions of model parameters. We use PyMC3 for sampling from the posteriors of the ESD parameters from different Galactic binary populations.

To perform the MCMC analysis, we utilise the PyMC3 default No-U-Turn sampling (NUTS) \citep[][]{2011arXiv1111.4246H} in our models. The NUTS algorithm is an extension of the Hamiltonian Monte Carlo (HMC) method used for sampling in Bayesian inference. Here are the steps involved in the NUTS algorithm:

\begin{enumerate}
    \item Initialisation: Start by initialising the Markov chain with a starting point $\theta$ in the prior parameter space $P(x)$.

    \item Initial Momentum Sampling: randomly sample momentum variables from a Gaussian distribution.

    \item Leapfrog Integration: Simulating the movement of the parameters in conjunction with their momentum variables according to the Hamiltonian dynamics.

    \item Stopping Criterion: The trajectory is continued until the "no-U-turn" condition, the point at which the trajectory "turns back" on itself, is met. 

    \item Adaptation and Splitting: If the no-U-turn condition is not met, the algorithm adapts by splitting the trajectory to explore multiple paths.

    \item Metropolis Acceptance: After the trajectory is determined, the proposed state is accepted or rejected probabilistically based on the Metropolis-Hastings acceptance criterion, which ensures the chain converges to the target distribution.

    \item Updating: The accepted state becomes the new starting point for the next iteration of the algorithm, and the process repeats until a sufficient number of iterations are reached.
\end{enumerate}

\section{Corner Plots}
\label{sec:BPASS_corner}
\begin{figure*}
    \centering
    \includegraphics[width=1\textwidth]{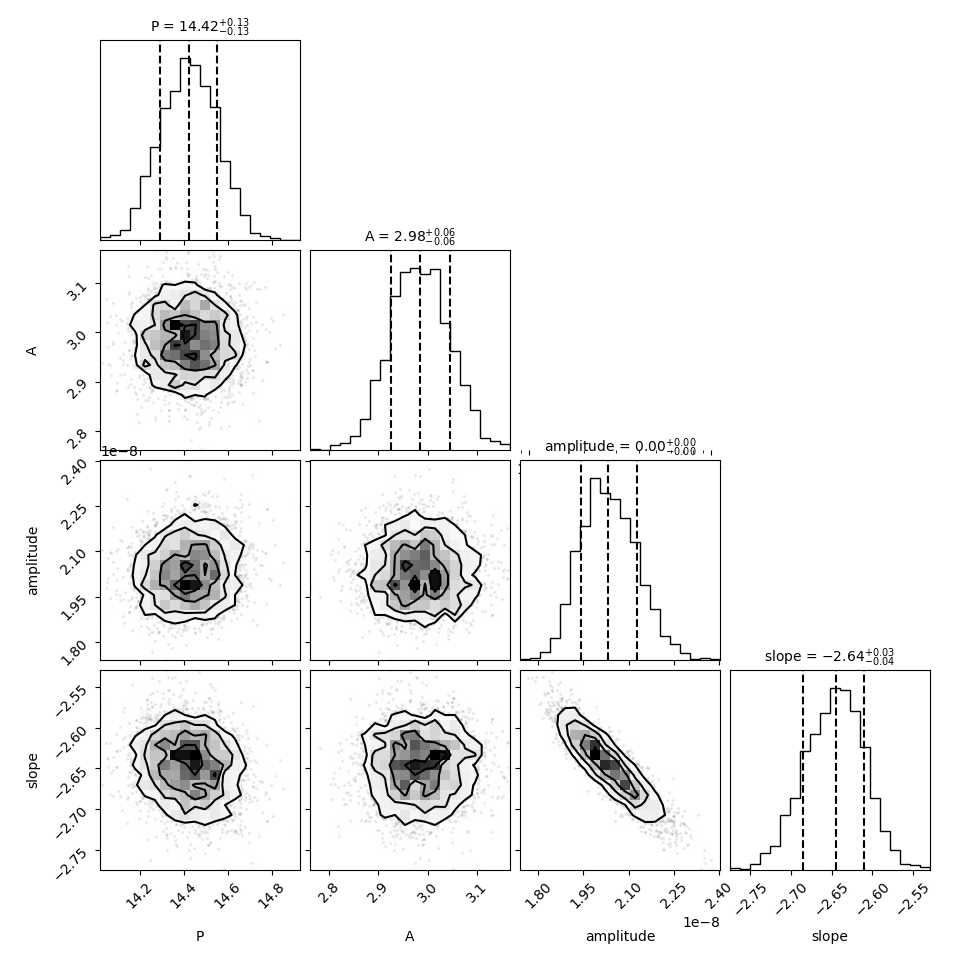}
    \caption{Corner plot for one MCMC estimation of four parameters. The results are for the LISA noise (parameters $P$ and $A$) and single PL (parameters $\alpha$ and $\beta$). The vertical dashed lines on the posterior distribution represent from left to right the quantiles [16 er cent, 50 per cent, 84 per cent].}
    \label{fig:pl_cornor}
\end{figure*}

\begin{figure*}
    \centering
    \includegraphics[width=\textwidth]{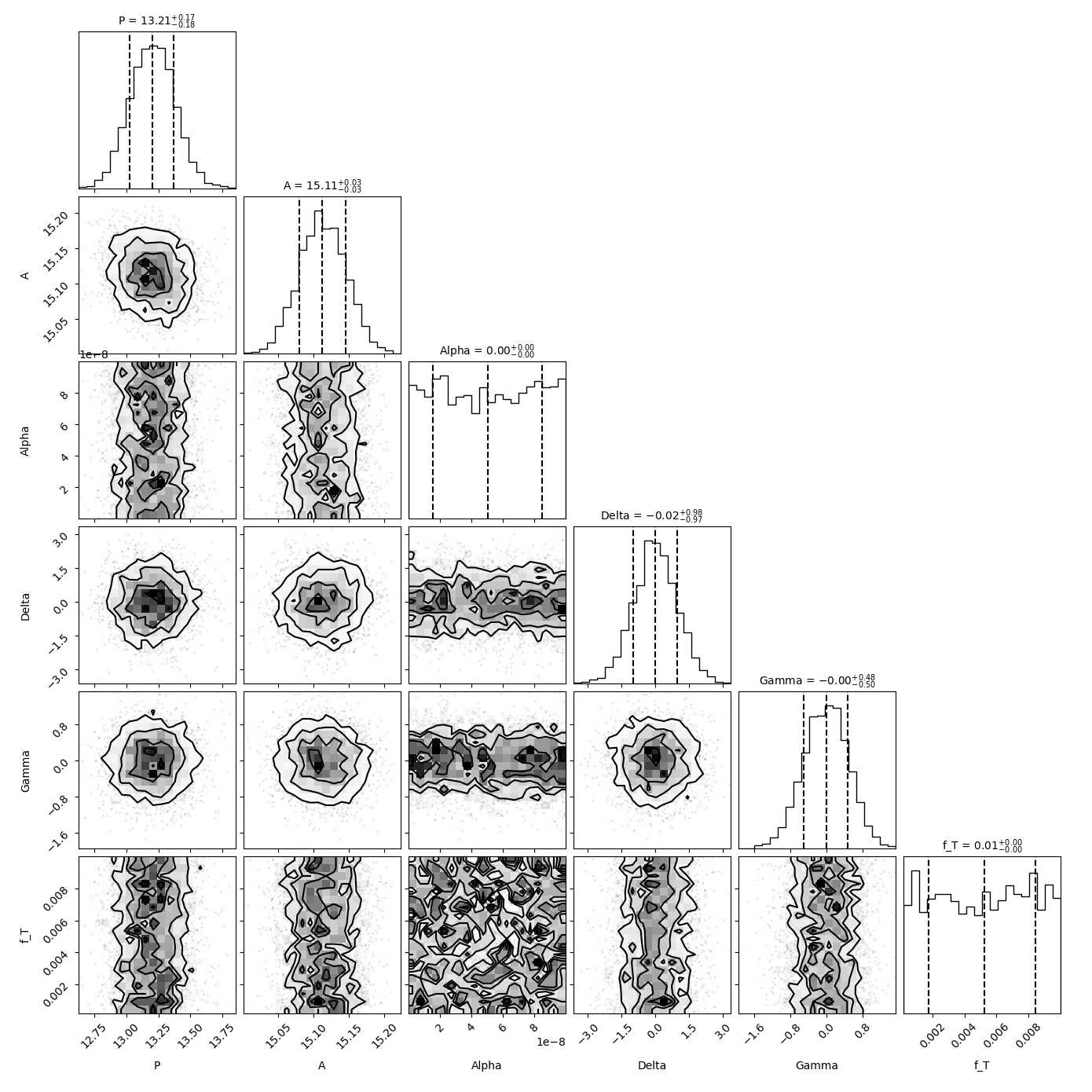}
    \caption{Corner plot for one MCMC estimation of five parameters. The results are for the LISA noise (parameters $P$ and $A$) and broken PL (parameters $\alpha$, $\delta$, $\gamma$ and $f_t$). The vertical dashed lines on the posterior distribution represent from left to right the quantiles [16 er cent, 50 per cent, 84 per cent].}
    \label{fig:broken_cornor}
\end{figure*}

\begin{figure*}
    \centering
    \includegraphics[width=\textwidth]{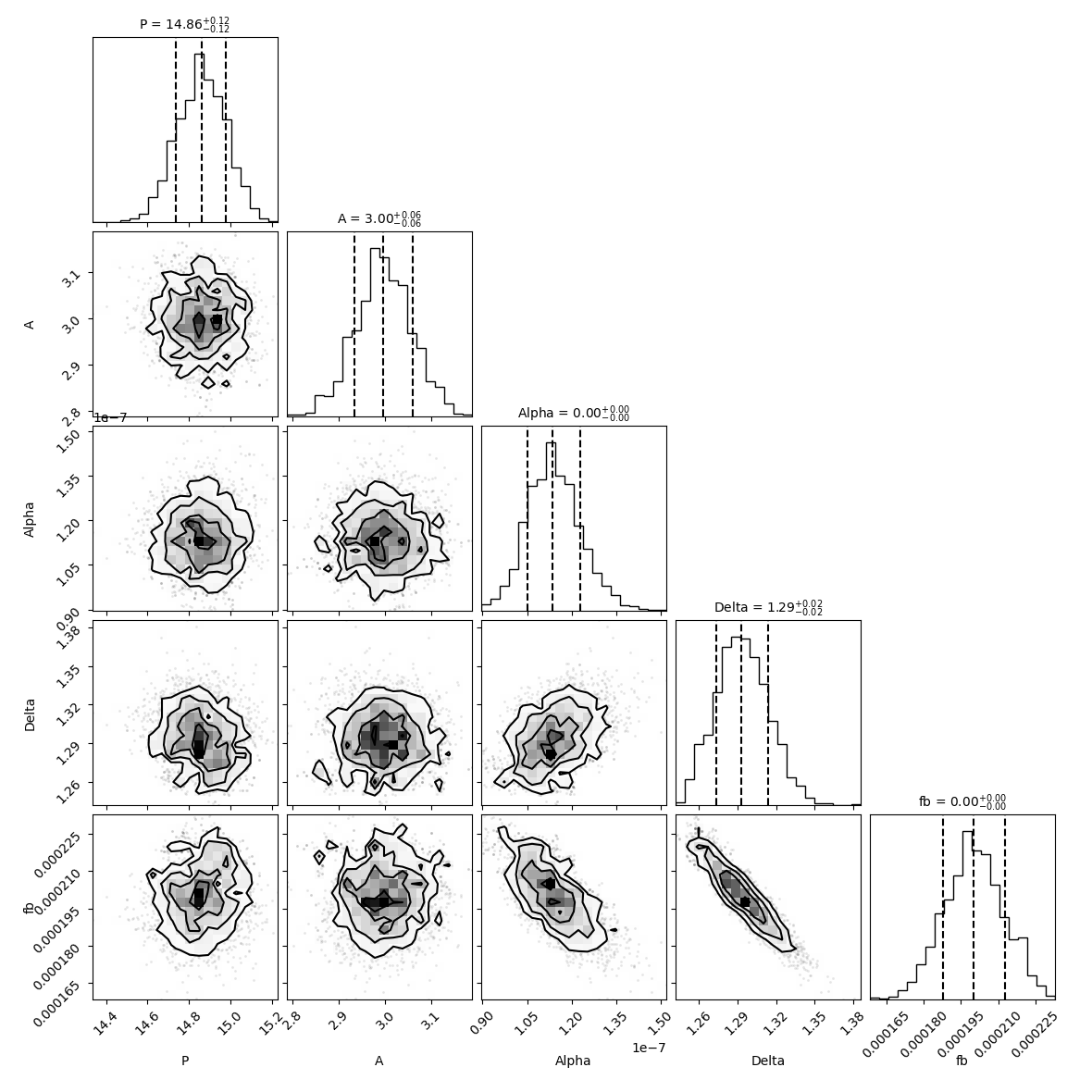}
    \caption{Corner plot for one MCMC estimation of five parameters. The results are for the LISA noise (parameters $P$ and $A$) and single-peak (parameters $\alpha$ $\Delta$ and $f_b$). The vertical dashed lines on the posterior distribution represent from left to right the quantiles [16 er cent, 50 per cent, 84 per cent].}
    \label{fig:peak_cornor}
\end{figure*}

One simulation result from one realisation with MCMC. We present the corner plots of all the parametric model estimates for the simulated signal in Figures~\ref{fig:pl_cornor}, \ref{fig:broken_cornor} and \ref{fig:peak_cornor}. The `true' noise parameters are $P=15$ and $A=3$ respectively. Both single PL and single-peak models noise parameters $P$ and $A$, and amplitude $\alpha$ converge well. However, the convergence fails in the broken PL model.

\section{Galactic ESD parameter estimates}

As supplementary information to Table~\ref{tab:5}, we present in Table~\ref{tab:6} the posterior summary statistics for all parameters of $h^2 \Omega_{\rm GW}$ for the broken PL and single-peak Bayesian models. The mean estimates are accompanied by the 90 per cent highest density interval (HDI). 
\renewcommand{\arraystretch}{2}
\begin{table*}
  \centering
  \caption{Posterior distributions for all the parameters in the broken PL and singel-peak Bayesian models. The mean estimates are accompanied with the highest density interval (HDI) is taken as 90 per cent. }
  \begin{tabular}{|c|c|c|c|c}
     & \multicolumn{2}{c}{\textbf{Broken PL}} \\
     \hline
        Population & $\alpha \times 10^{-12}$   & $\delta$                     & $\gamma$                  \\
        \hline
        WDB        & $500^{+32}_{-41}$          & $-0.01 ^{+1}_{-1}$           & $-0.01 ^{+1}_{-1}$      \\
        BHB        & $5^{+4}_{-4}$              & $-0.01 ^{+1.05}_{-1.02}$     & $0.00 ^{+1.05}_{-1.04}$  \\
        NSB        & $5^{+4}_{-4}$              & $-0.01 ^{+1.0}_{-0.96}$      & $0.04 ^{+1.00}_{-1.04}$  \\
        NSWD       & $5^{+4}_{-4}$              & $-0.01 ^{-+0.99}_{-1.01}$    & $-0.04 ^{+1.04}_{-0.98}$ \\
        BHNS       & $5^{+4}_{-4}$              & $-0.01 ^{+1.0}_{-1.0}$       & $-0.01 ^{+1.0}_{-1.0}$   \\
        BHWD       & $500^{+400}_{-400}$        & $-0.01 ^{+0.98}_{-1.04}$     & $-0.01 ^{+0.98}_{-0.99}$ \\
    \hline
     & \multicolumn{2}{c}{\textbf{Single-peak}} \\
     \hline
        Population       & $\alpha \times 10^{-12}$        & $\delta$                  & $f_b \times 10^{-5}$      \\
        \hline
        WDB              & $610^{+80}_{-60}$               & $1.71 ^{+0.04}_{-0.04}$   & $0.00 ^{+0.00}_{-0.00}$  \\
        BHB              & $0.45^{+0.4}_{-0.3}$            & $5.2 ^{+3.17}_{-3.26}$    & $0.00 ^{+0.00}_{-0.00}$  \\
        NSB              & $1.4^{+0.4}_{-0.4}$             & $8.03 ^{+4.59}_{-4.68}$   & $0.00 ^{+0.00}_{-0.00}$  \\
        NSWD             & $1.3^{+0.2}_{-0.2}$             & $8.31 ^{+5.23}_{-5.20}$   & $0.00 ^{+0.00}_{-0.00}$  \\
        BHNS             & $0.45^{+0.4}_{-0.3}$            & $7.88 ^{+4.98}_{-5.02}$   & $0.00 ^{+0.00}_{-0.00}$  \\
        BHWD             & $500^{+140}_{-200}$             & $1.05 ^{+0.06}_{-0.04}$   & $0.00 ^{+0.00}_{-0.00}$  \\
    \label{tab:6}
  \end{tabular}
\end{table*}


\bsp	
\label{lastpage}
\end{document}